\documentclass{aastex}
\usepackage{emulateapj5}
\usepackage{graphicx}
\usepackage{psfig}
\slugcomment{To appear in the Astrophysical Journal, March 2001}
\begin{document}
 
\title{NICMOS Imaging of the Host Galaxies of $z \sim 2$ -- 3 Radio-Quiet
Quasars\altaffilmark{1}}
\altaffiltext{1}{Based
on observations made with the NASA/ESA Hubble Space Telescope, obtained
at the Space Telescope Science Institute, which is operated by the
Association of Universities for Research in Astronomy, Inc., under
NASA contract NAS 5-26555.}

\shorttitle{NICMOS Imaging of $z \sim 2$ -- 3 RQQ Hosts}

\author{Susan E. Ridgway and Timothy M. Heckman}
\affil{Johns Hopkins University, Dept. of Physics and Astronomy, 
3400 N. Charles St. Baltimore MD 21218\\E-mail:ridgway@pha.jhu.edu and 
heckman@pha.jhu.edu}
\author{Daniela Calzetti}
\affil{Space Telescope Science Institute, 3700 San Martin Dr., Baltimore,
MD 21218\\E-mail:calzetti@stsci.edu}
\author{Matthew Lehnert}
\affil{MPE, Postfach 1603, D-85740 Garching, Germany\\E-mail:mlehnert@qso.mpe-garching.mpg.de}

\begin{abstract}
\tighten

We have made a deep NICMOS imaging study of a sample of 5 $z \sim 2$ -- 3
radio-quiet quasars with low absolute nuclear luminosities, and
we have detected apparent host galaxies in all of these.
Most of the hosts have luminosities approximately equal to
present-day L$_*$, with a range from 0.2 L$_*$ to about 4 L$_*$.
These host galaxies have magnitudes and
sizes consistent with those of the Lyman break galaxies
at similar redshifts and at similar rest wavelengths, but
are about two magnitudes fainter than high-$z$ powerful
radio galaxies.
The hosts of our high-$z$ sample are comparable to or less luminous than 
the hosts of the low-$z$ radio-quiet quasars with similar 
nuclear absolute magnitudes. 
However, the high $z$ galaxies are
more compact than the hosts of the low $z$ quasars, and probably have
only 10 -- 20\% of the stellar mass of their low-$z$ counterparts.
In one host, we find a residual component that is not centered on the quasar
nucleus, and several hosts have apparent companions within a projected
distance of $\sim$10 kpc, 
indications that these systems are possibly in some phase of a
merger process.
Application of the $M_{bulge}$/$M_{BH}$ relation found for present-day 
spheroids to the stellar masses implied for the high $z$ host galaxies
would indicate that they contain black holes with masses around $10^{8}$
$M_\odot$. Comparison to their nuclear magnitudes implies accretion
rates that are near or at the Eddington limit. 
Although these high $z$ hosts already contain supermassive 
black holes, the galaxies will need to grow significantly to evolve
into present-day $L_*$ galaxies.
These results are basically consistent 
with theoretical predictions for the hierarchical buildup of the galaxy host
and its relation to
the central supermassive black hole. 

\end{abstract}

\section{Introduction}

Radio-quiet quasars (RQQ) represent the great majority of the known active
galactic nuclei (AGN) at
high redshift ($z \gtrsim 1$), yet
information on the galaxies that host these
AGN is currently extremely limited. 
At lower redshift, luminous RQQ are found in 
both spiral and elliptical hosts, with the
most luminous radio-quiets appearing in the bulge-dominated systems.
The host galaxy luminosities correspond to
a few times a present-day $L_*$ galaxy (e.g. Bahcall et al. 1997;
McLure et al. 1999).
In addition, the
luminosity of the host seems to correlate roughly with that of the nucleus
(McLeod \& Rieke 1995; Hooper et al. 1997;
McLure et al. 1999; McLeod, Rieke, \& Storrie-Lombardi 1999).
Nearby, almost
every bulge-dominated galaxy may contain a supermassive
black hole candidate whose mass is roughly proportional to the luminosity of its
bulge (Kormendy \& Richstone 1996; Magorrian et al. 1998; van der Marel 1999). 
The black hole masses are 
even better correlated with the stellar velocity
dispersions observed in the bulges (Ferrarese \& Merritt 2000;
Gebhardt et al. 2000a). These relationships imply a close correlation
between the mass of the stellar bulges in these systems and the masses of
their resident black holes. 

The evolution in the star formation rate from
an epoch of $z \sim 2$ to the present day may be qualitatively similar to
the huge drop
in the number density of quasars from the peak at $z \sim$ 2 -- 3 to the
present epoch, which is roughly similar to the
evolution in the comoving number density of
radio sources (e.g, Dunlop 1998; Boyle \& Terlevich 1997).
All of these factors suggest a 
strong link between the formation and evolution
of galaxies and those of the quasars and their hosts, and support
the idea that an active quasar may be a short-lived episode in the
life of every bulge-dominated galaxy.

Kauffmann \& Haehnelt (2000) have investigated the evolution of the hosts of
the radio-quiet quasars
and the formation and fuelling of their central black holes using
semi-analytic hierarchical clustering models of
galaxy formation, and have made specific predictions 
regarding the evolution of this nuclear magnitude--host galaxy luminosity
relation. The high-$z$ radio-quiet quasars are far less
studied than the radio-loud host galaxy population, and few observations
have been made at nuclear luminosities similar to those of 
the low-$z$ radio-quiet sample.

To study the evolution of the radio-quiet quasar hosts and their
relationship to the other samples of high-$z$ galaxies known,
we have made HST NICMOS observations of radio-quiet quasar hosts near the
epoch of the peak quasar density ($z \sim$2 -- 3). We have chosen
a sample of quasars whose nuclei
are faint enough to provide a good comparison sample
to those of the well-studied low-$z$ quasars.
Throughout this paper we use  $H_0$ = 50 km s$^{-1}$ Mpc$^{-1}$
and $\Omega$  = 1, unless otherwise stated.

\section{Observations and Data Reduction}

\subsection{Sample Selection}
We have selected 5 quasars with B$>$21.5 at $z \sim$ 2--3 
from the faint quasar sample of Zitelli et al. (1992), which is
a spectroscopically-complete subsample from the deep optical quasar survey
of Marano, Zamorani \& Zitelli (1988).
The exact redshift ranges were constrained
to avoid strong emission lines falling in the NICMOS filter bandpasses
near 1.6 $\mu$m; this
resulted in a sample of 3 objects at $z \sim$1.8 and 2 objects at $z \sim$2.7.
Their nuclear magnitudes are in the range $\rm M_{\rm B}$ = 
$-$22 to  $-24$, making them comparable 
in absolute magnitude to many 
low-$z$ quasar samples (e.g. Bahcall et al. 1997; McLure et al. 1999; McLeod
et al. 1999).
One object, MZZ 9592, is a BAL QSO (Zitelli et al. 1992).

\subsection{Observations}

The observations were made using the NIC2 aperture of HST's NICMOS
camera, which has a field size of 19\farcs2 $\times$ 19\farcs2,
at a scale of 0\farcs075 pixel$^{-1}$.
To achieve emission-line-free imaging, the $z \sim 1.8$ objects were imaged
in the F165M filter, resulting in a rest-frame wavelength of $\sim$V,
and the $z \sim 2.7$ objects in the F160W filter, corresponding to 
rest-frame $\sim$B. 
To convert count rates into fluxes in Jy, we have used for each filter
the HST-supplied calibration factors, i.e. 1 count s$^{-1}$ is 2.07 $\mu$Jy
for F160W and 4.07 $\mu$Jy for F165M. 
Where we have used $B$, $V$, $R$, or $H$-band magnitudes, 
we have used zero points of 4260, 3540, 2870 and 980 Jy respectively. 

We observed each of the 5 quasars (and its corresponding PSF star) with two
visits separated by several
months, resulting in observations of each field at significantly
different position angles on the sky relative to the PSF pattern
(which is fixed in the spacecraft frame).  This allows an independent check
on the reality of any residual emission seen. 
We used a dither pattern with offsets of $\sim$3\arcsec (40 pixels) between
integrations, for 6 or 8 integrations per visit. The dither pattern
(suggested to us by M. Dickinson, priv. comm. 1997) 
included half-pixel offsets to improve resolution
by adequately sampling the HST PSF at these wavelengths.
The final FWHMs achieved were 0\farcs14 -- 0\farcs16.
We chose a nearby star for each of the 5 quasars, and observed this 
star in the same visit as the quasar using the identical dither pattern
in order to characterize the point spread function (PSF). 
We imaged in the MULTIACCUM read mode with (primarily) the STEP128 
sample sequence for the quasar integrations, resulting in  
observing times per frame of $\sim$ 900 -- 1500 s for the quasars, 
and 6 -- 15 s for their PSF stars.
Total exposure times for the quasars were
6000 -- 10000 s per visit.
However, due to variable
cosmic ray persistence problems induced by SAA passages in some
of the exposures, the amount of useful data 
varied from visit to visit. 
The sky noise levels in the final images therefore differ.
We give the summed integration times of data actually used in the final
images in the log of observations in Table \ref{obslog}.  

\subsection{Calibration}

We recalibrated the raw NICMOS data using a modified version of
the standard pipeline process. 
After identifying the most up-to-date
calibration files, we used the STSDAS task {\it calnica}
in a first pass to make basic corrections for the bias and dark contributions,
and the non-linearity of the device. We also 
applied the ``CRIDCALC'' algorithm which
adds the multiple reads and removes the worst of
the cosmic rays. 

The primary bias anomalies we encountered in our data were the readout artifacts
termed ``bars'', the bias gradients termed ``shading''  and 
the variable quadrant bias called the ``pedestal effect'', 
described in Skinner et al. (1998) and the
NICMOS Data Handbook v. 4.0 (December 1999).
We masked by hand any bars that were not removed by the {\it calnica} routine.
To remove the quadrant-dependent pedestal 
from the frames before flat-fielding,
we used the technique developed by Mark Dickinson:
his script {\it pedsky} and the non-interactive version of {\it pedsky} 
called {\it azped2} written by Andrew Zirm and Mark Dickinson.
This method 
works by determining the best value for 
the sky signal plus quadrant-dependent constant offsets (pedestals) 
by subtracting various scaled versions of the flatfield, removing 
an estimated residual bias for each quadrant in this subtracted image,
filtering out objects with a median smoothing, then calculating the 
RMS for the image. The sky value and pedestal values
that yield the minimum noise are adopted. 
This method works well at improving the final flatfielded images for our
fairly empty fields.
The vertical bands caused by the bias ``shading'' effect 
were removed by using the IRAF routine {\it background} to 
fit constant values to the columns.

\subsection{Removal of persistent cosmic rays}
 
The resultant images still vary greatly in quality and depth; this is 
primarily the result of cosmic ray (CR) persistence after a 
passage through the South Atlantic Anomaly (SAA). The nature of this
problem is more fully discussed in Najita, Dickinson \& Holfeltz (1998).
Cosmic ray hits during these passages leave residual signal on the detector
that decays with time. In our data, 
depending on the depth of the SAA passage, this 
residual noise signal can quadruple the effective sky noise and greatly
affect the usefulness of specific frames. 
In many cases, half or more of the sky area is affected by 
persistent ghosts of CRs.
In very severe examples, the frames were saturated by the 
CR bombardment, and the residual signal takes on a characteristic
large scale shape which is correlated to detector properties
rather than to the actual location of the hits. In such images, specific
spatial masking of the CRs is impossible. 

Visits in which CR persistence was a severe problem generally
had observations taken in 
pairs after each SAA passage, with the first much 
more heavily affected by the CR persistence, and the second carrying the 
same signal, but at a faded, lower intensity. 
We can therefore use the frames which have the strongest signature of 
the SAA passage 
to remove some of the faded CR signal from the subsequent frame, and therefore
greatly decrease the artificial non-Gaussian noise in those frames. 
We created an object mask from
a combined frame, and used this object mask to mask out the objects in the badly
impacted frames, leaving the background. We then subtracted this 
background-only frame (multiplied by the 
appropriate scaling factor) from the subsequent frame to remove the persistent
CR signal. 
We used a routine written by Eddie Bergeron to determine the best scaling
factor by minimizing the noise in the sky histogram of the difference frame. 
In many frames this resulted in a 20 -- 40\% decrease in the effective
noise, and resulted in much flatter frames. We then excluded
the severely-impacted frames used to model the CR-persistent pattern
from the final combination of data. 

In general, 
PSF star observations and object observations were calibrated identically.
Of course, because of the relatively short exposure times,
CR persistence was not a problem for the PSF star observations.

\subsection{Combination of the data}

Since we have two visits to each object which are at different
position angles on the sky,
we combine and PSF-subtract each of the two visits to each object
separately. This allows us to compare two independent determinations of 
the fluxes and positions of features in the residual hosts. 
To determine the relative linear shifts between dither positions, 
we have calculated centering both from the
objects on the frames and from the 
spacecraft jitter files (where this information
was complete), and found no significant difference in result. 
We then combined the frames, using a simple method
that uses the STSDAS routine CRREJ to determine the locations of the CR 
affected pixels in the initial
frames and creates bad pixel masks, resamples the corrected frames
to double the linear dimensions, and combines these using standard rejection
techniques. 
The visits that are most affected by the SAA passages may have the number of
individual frames reduced to as few as 3, giving the combined
image a resolution $\sim$10\% worse
Use of the ``drizzle'' technique to shrink the output pixels did not
significantly improve the final resolution for our data, probably
because we had a limited number of sky positions and the HST PSF at 1.6$\mu$m
is not severely undersampled by NIC2.

\section{Data Analysis}

\subsection{PSF subtraction}

We have made a simple, direct two-dimensional
PSF subtraction for each visit on a quasar
through an iterative method. 
We fit a flat background in the combined frame
by calculating the mean sky value in an annulus around the quasar that excludes
nearby companion objects and most extended flux (generally from about 1\farcs5
to 3\arcsec).
We centroid the PSF and quasar within a central radius
of 0\farcs09, and scale the PSF flux to the quasar flux in this aperture.
We then vary the centering of the PSF, subtract the two images, and
minimize the $\chi^{2}$ of the residuals to determine the best center. 
Finding the best relative center is unambiguous 
since off-center subtractions are obvious in the residuals, and
we can treat the scaling and centering independently. 
Finding the best scaling is more subjective. 
We would like the residual ``host galaxy''
to have properties like a real galaxy, and at the least to have a radial profile
that increases monotonically. 
The simplest way of ensuring this in two dimensions is
to require flatness of the residual across some central region.
We vary the PSF-quasar scaling of the 
subtraction until the mean value in the residual of an annulus (from 0\farcs09
-- 0\farcs15) 
is equal to that of the inner 0\farcs09 aperture. 
This method is discussed further
in Ridgway \& Stockton (1997). 
This will still likely result in an oversubtraction, depending
on how peaked the real host galaxy is within this inner radius.
If there is no significant excess 
flux in the outer annulus, the PSF-quasar scaling determined
will be equivalent to the most conservative limit of subtracting
the central region to zero residual flux. 

We achieved the cleanest subtraction with the PSF star observed during the
same visit as the quasar; for one visit of MZZ 1558, we were unable to 
obtain a PSF star, and we have used the PSF star from the other visit
without major problems.
The count rates in the observed PSF stars are in all cases thousands of times 
greater than those of the quasars and
therefore signal-to-noise in the PSFs was not a concern. The scaling factors
(QSO/PSF) range from 10$^{-3}$ to 10$^{-5}$ for the PSF
stars; the MZZ9592 quasar-to-field-star ratio is 0.26. 
The PSF scaling factors differed by $<$1\% up to 8\% between the two visits
for each object.

As a result of this subtraction process, we find resolved excess 
two-dimensional flux around the quasars relative to the PSF stars, in 
each of the two visits to each quasar.
The quasar hosts are also resolved relative to a star with an apparent magnitude
similar to those of the quasars that fell within the field of MZZ 9592.
This field star provides a good check that our results are not the product of
some difference in the way we have observed and reduced the PSF stars
versus the quasars. We have thus also treated this star as if it were a
quasar and applied the same PSF subtraction techniques to it; we find no
significant flux in the residual.
 
\begin{figure}
\begin{center}
\begin{picture}(350,450)
\put(-150,-160){\includegraphics{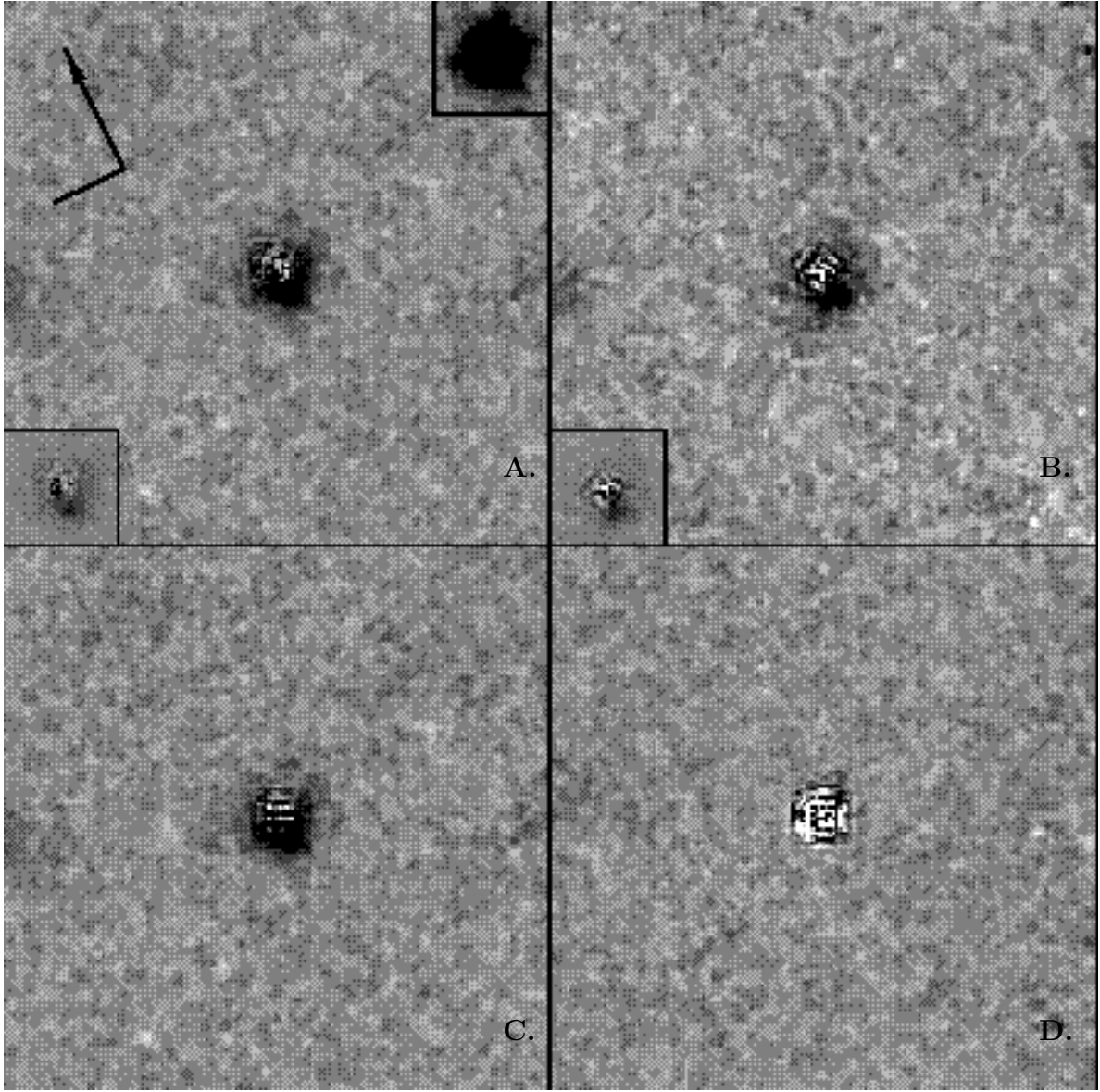}}
\put(138,265){\large{\bf A.}}
\put(357,265){\large{\bf B.}}
\put(138,35){\large{\bf C.}}
\put(357,35){\large{\bf D.}}
\end{picture}
\end{center}
\noindent
\caption{MZZ 9592, PSF subtractions and tests. All frames are 5\farcs7 square;
the sky position angle is indicated by the compass arrow to north. 
{\it A.} Visit one, PSF-subtracted,
using the observed bright PSF star from the same visit.
The central region of this same image
(at a lower display stretch) is shown in the lower left inset,
showing an off-center residual host component.
The upper right inset shows the unsubtracted quasar.
{\it B.} Visit two, PSF-subtracted, rotated to match the orientation of
visit one. Sky noise is worse in this image, with residual low-level CR 
problems, but the off-nuclear component is visible (also seen in
the lower left inset of the central region). 
{\it C.} Visit one, PSF-subtracted using the star that falls within the
field, demonstrating that the residual is not an artifact
of the mismatch between the observational
strategies applied to the quasar fields and the bright PSF stars.
{\it D.} The field star minus the PSF star: no net residual flux. 
\label{m9592testpsffig}
}
\end{figure}

In Figure \ref{m9592testpsffig}, 
we show the results of these analyses for both visits
to MZZ 9592, separately. 
The extension
around MZZ 9592 consists of a bright, relatively symmetric 
halo of emission relative to both the PSF and the field star,
but has in addition an off-nuclear component which will allow
us to check that our resolved morphology is stable. 
This component appears at the same location in both visits
and regardless of whether the field
star or the observed PSF star is used for subtraction. 
Results of aperture photometry of the residual are consistent
within a few percent
for each visit and subtraction variation. These tests show that our results
are not likely to be an artifact of our observing or analysis method. 

\subsection{PSF-subtracted host galaxy images}

To create the best S:N image of the extended host,  
we combine the PSF-subtracted images from the
two visits by centering using the quasar,
rotating using the nominal sky position angle (ORIENT header parameter),
and averaging, after weighting the two visits by the inverse 
variance of the sky noise. 

\begin{figure}
\begin{center}
\begin{picture}(350,550)
\put(-130,-80){\includegraphics{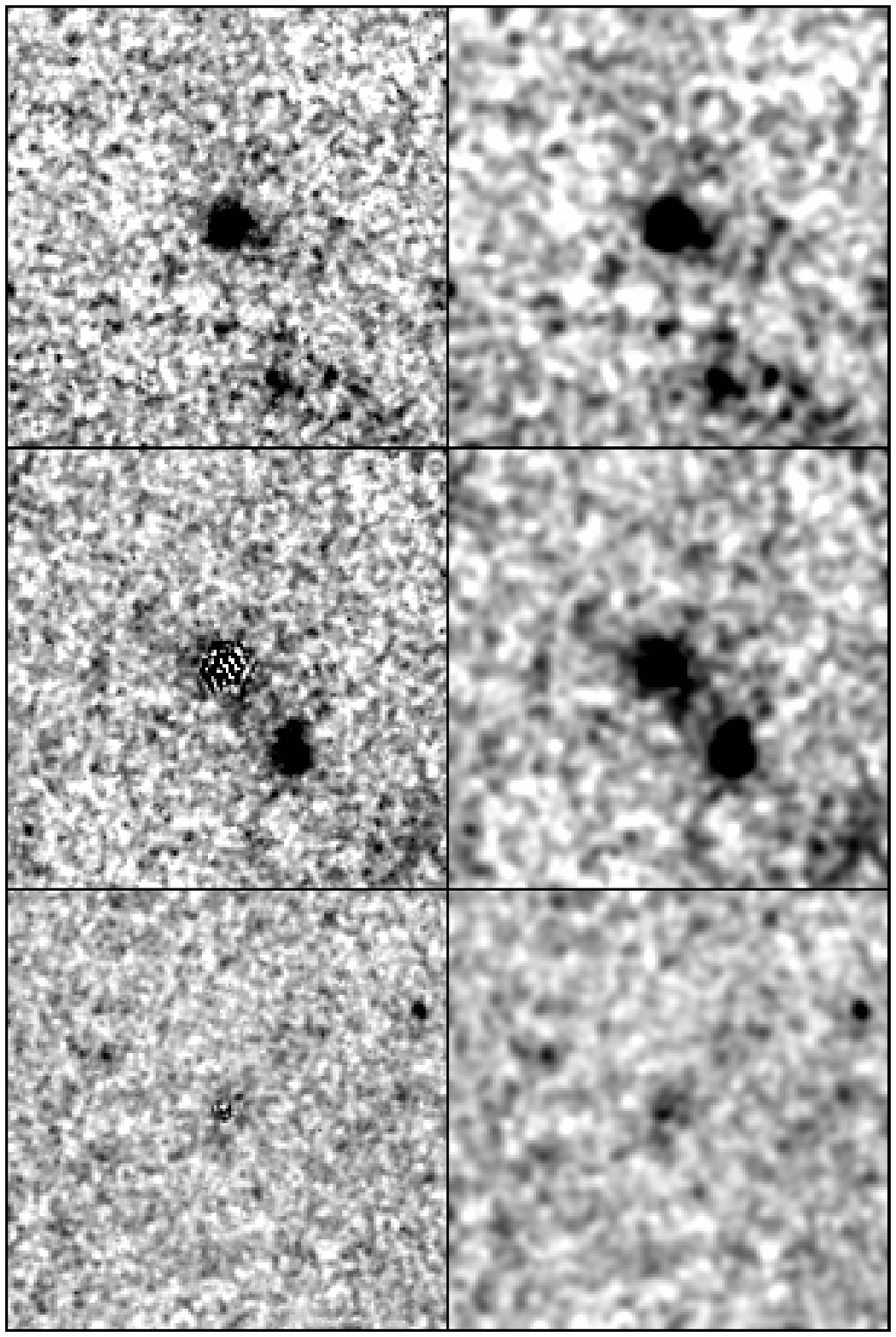}}
\put(-90,455){\large{\bf MZZ 11408}}
\put(60,550){\large{\bf Raw}}
\put(210,550){\large{\bf Smoothed}}
\put(-90,260){\large{\bf MZZ 1558}}
\put(-90,120){\large{\bf MZZ 4935 }}
\end{picture}
\end{center}
\noindent
\caption{PSF-subtracted $z \sim 2$ MZZ quasar hosts. Right panels
show images that are smoothed
with a Gaussian kernel with $\sigma$$=$0\farcs06.
Each panel is 5\farcs7 square (or roughly 45 kpc), N up, E left.
\label{z2mzzfig}
}
\end{figure}

\begin{figure}[p]
\begin{center}
\begin{picture}(350,350)
\put(-130,-100){\includegraphics{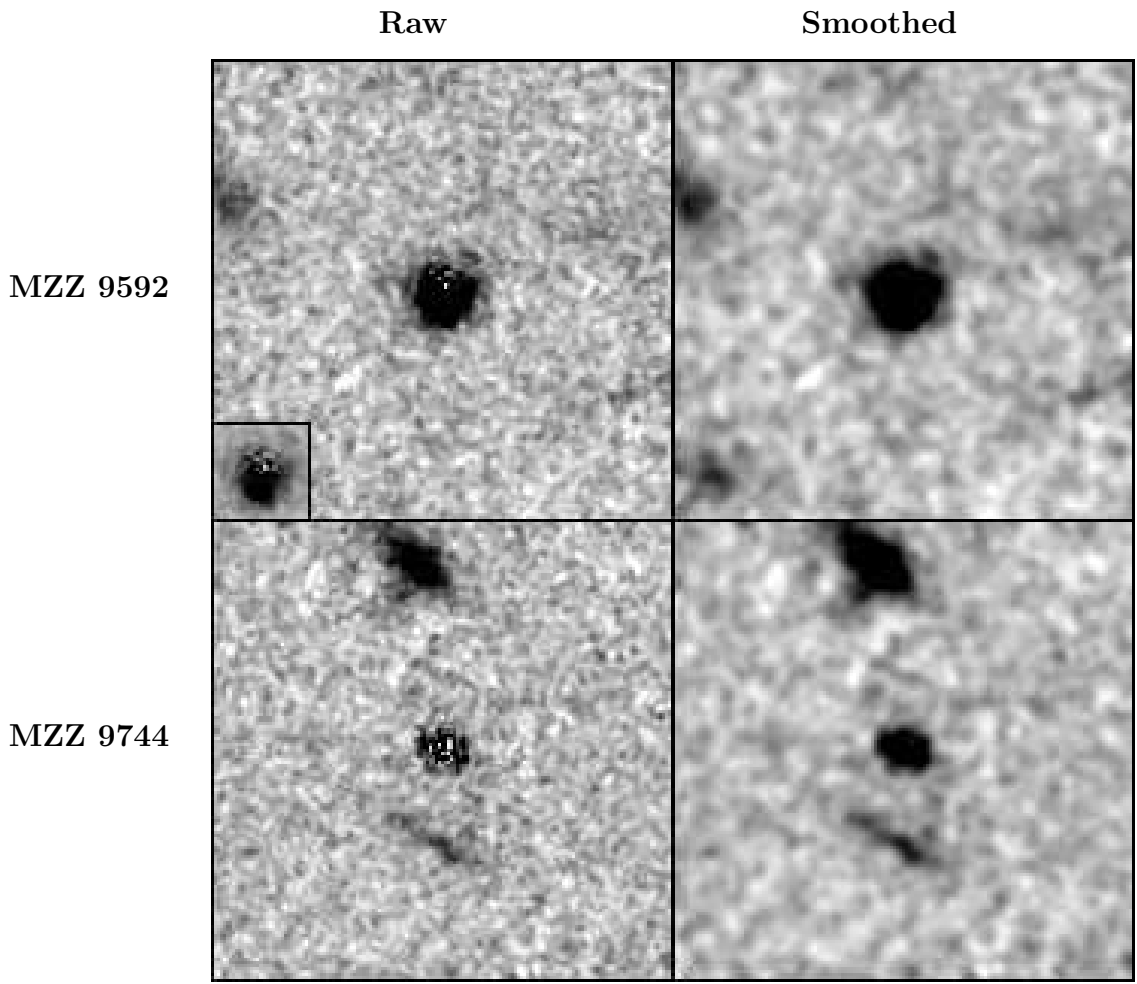}}
\put(55,360){\large{\bf Raw}}
\put(215,360){\large{\bf Smoothed}}
\put(-85,260){\large{\bf MZZ 9592}}
\put(-85,90){\large{\bf MZZ 9744}}
\end{picture}
\end{center}
\noindent
\caption{PSF-subtracted $z \sim 3$ MZZ quasar hosts, on the left unsmoothed;
on the right, smoothed
with a Gaussian kernel with $\sigma = $0\farcs06 . Each panel is 5\farcs7 square
(or roughly 45 kpc), N up, E left.
MZZ 9592, left panel has an inset of the central region at less
display stretch, showing the off-center host component.
\label{z3mzzfig}
}
\end{figure}

In Figures \ref{z2mzzfig} and \ref{z3mzzfig},
we give these PSF-subtracted combined results
for the 5 quasars, both at the original resolution and smoothed
with a Gaussian kernel with $\sigma =$ 0\farcs06. 
In each of the five quasars we detect extended flux around the 
quasar nucleus. 
The faintest residual is that of 
MZZ 4935, whose extended residual we detect at about the 3$\sigma$ level. 

The PSF-residuals create the noisy, ringing patterns seen in the
central regions of the images, with a radial extent that varies
slightly depending on the brightness of the quasar nucleus and details of
the PSF observations. The noisy regions are mostly $<$0\farcs15 in radius.
For MZZ 1558, however, we did not have a PSF star from the same visit for
one of our observations, and this PSF--quasar fit is not as good,
resulting in obvious subtraction residuals to a larger radial extent.
In all cases, however, we detect flux outside of these noisy inner
regions. 

\subsection{One-dimensional enclosed flux and radial profile plots}

\begin{figure}
\plotone{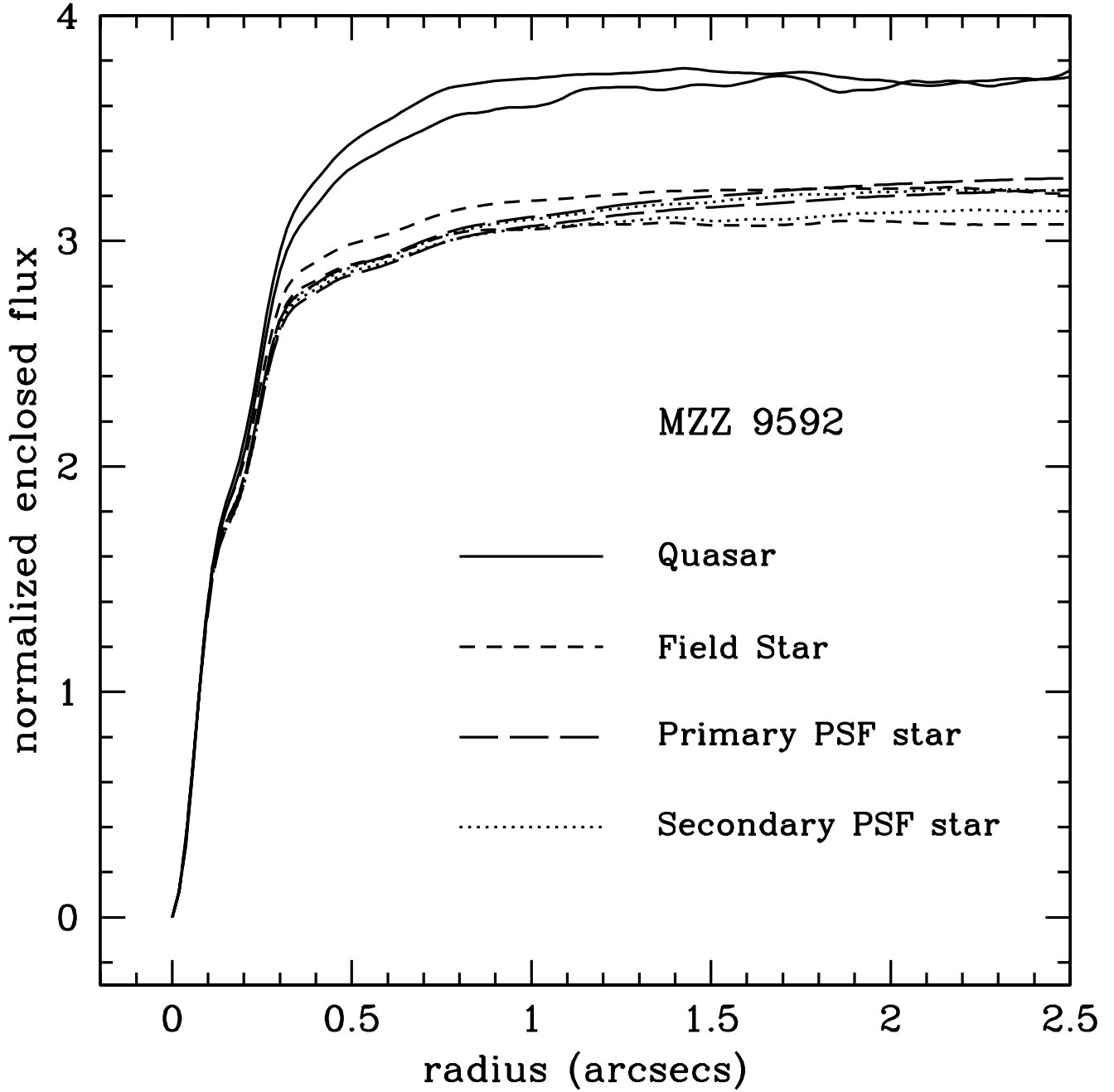}
\caption{The enclosed flux is shown versus the radius of the
aperture for all observations of MZZ 9592. The solid lines are the
quasar in each of the two visits, while the dashed and 
dotted lines represent
various stars observed: the field star which was observed simultaneously
with the quasar, the primary PSF star, and a secondary PSF star of reasonable
brightness which fell on the PSF star field. The enclosed flux has been
normalized in all cases to 1 for the aperture with radius 0\farcs075, 
corresponding to a diameter of $\approx$ 1 FWHM.
\label{m9592profilefig} }
\end{figure}

Another direct way of presenting the difference between the PSF stars
and the quasars is by comparing radial 
plots of the enclosed cumulative flux, providing
a one-dimensional, azimuthally-averaged view of the excess flux.
We show in Figure \ref{m9592profilefig} the results of making this
analysis for both observations of the quasar MZZ 9592, and for
both observations
of the available comparison stars: the primary bright
PSF star, a secondary bright PSF which fell off-center but within the
PSF star field, and the field star of comparable brightness to the 
quasar which fell on the quasar field. The enclosed flux at each radius
is normalized to that found within a radius of 0\farcs075, which 
corresponds to a diameter of $\approx$ 1 FWHM. The two quasar
enclosed flux profiles deviate significantly from all of the
stellar enclosed flux profiles, despite some variation in both quasar
and PSF profile from observation to observation. 

\begin{figure}[p]
\begin{picture}(400,500)
\put(25,205){\includegraphics{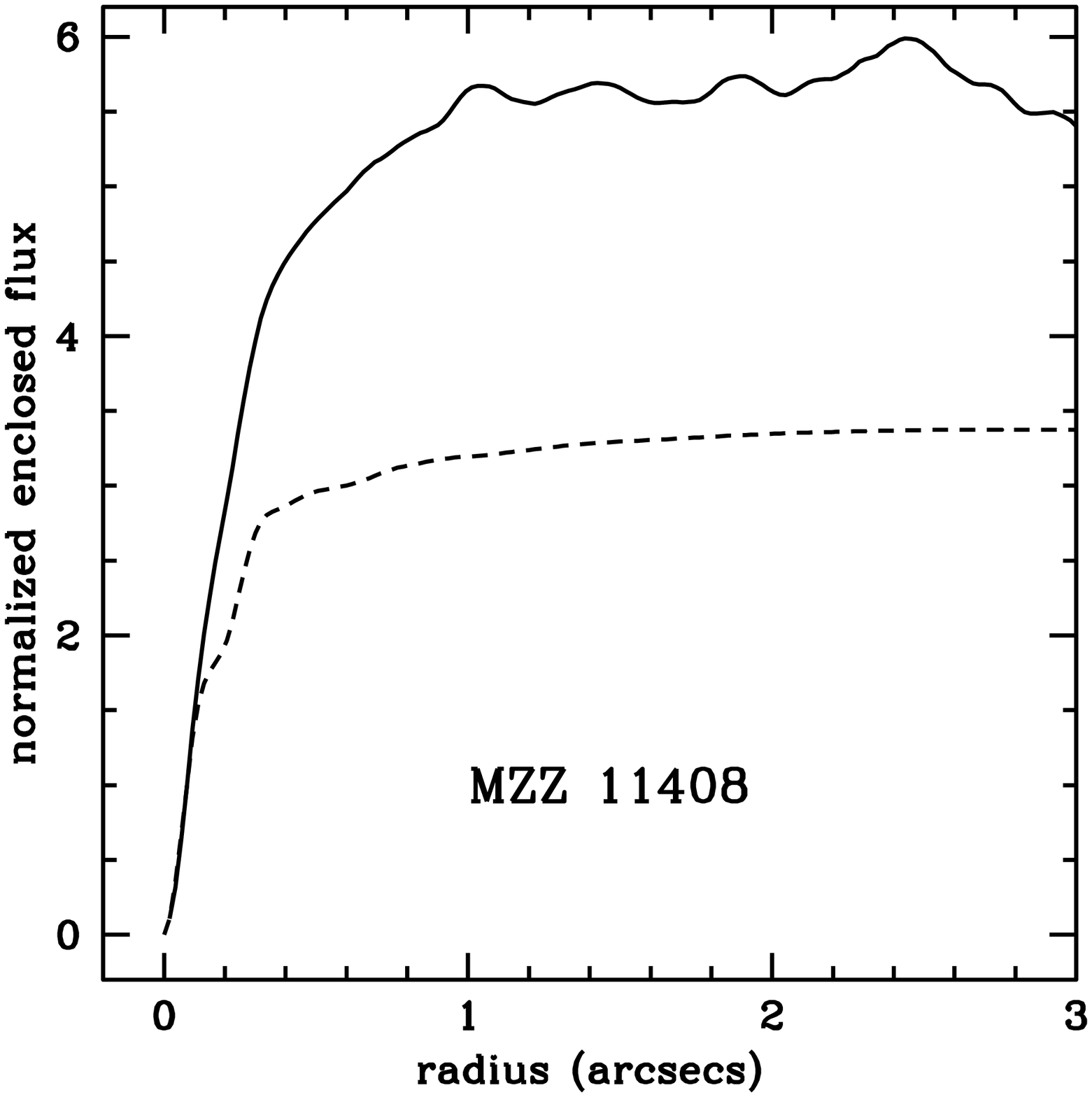}}
\put(255,205){\includegraphics{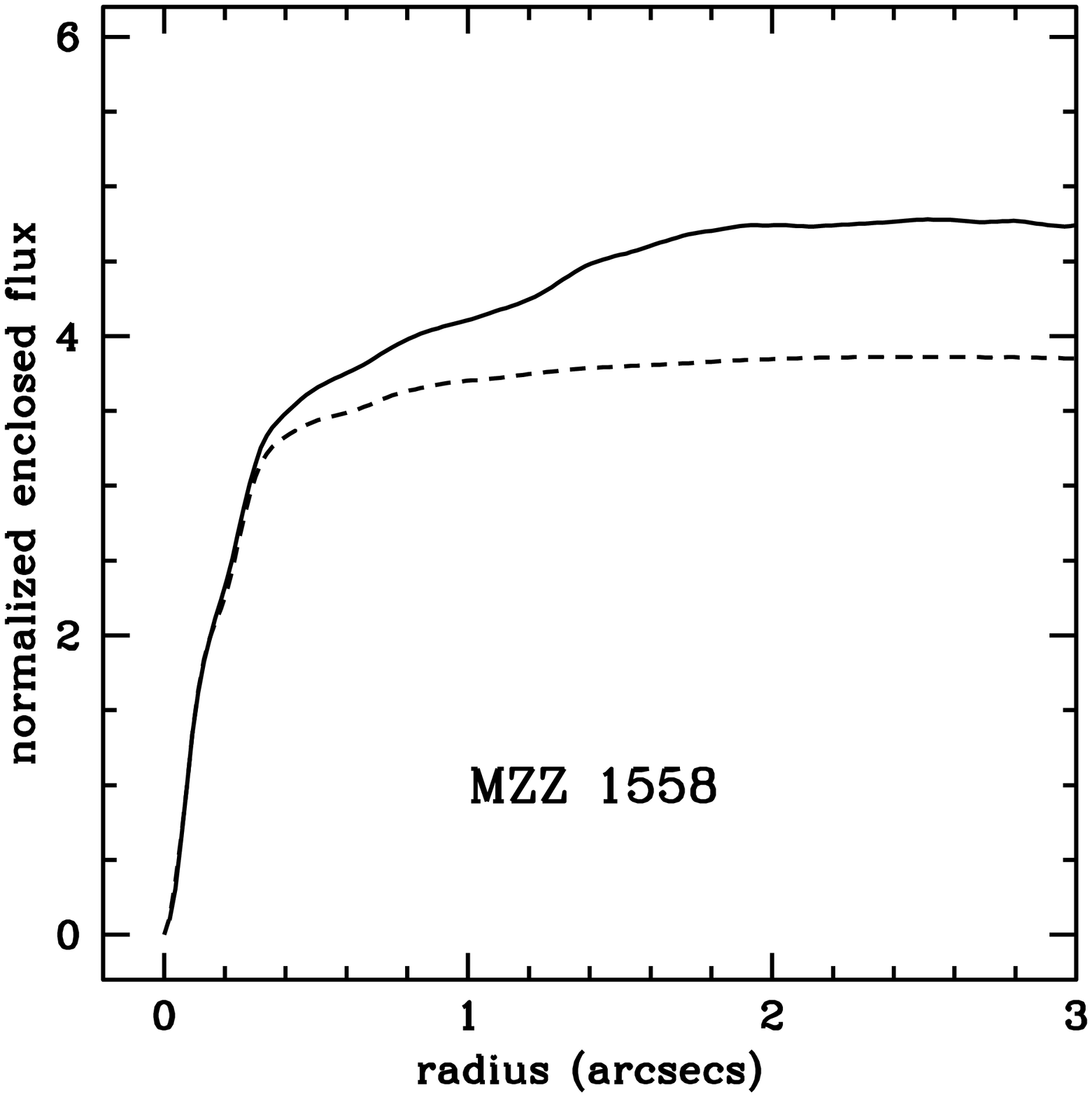}}
\put(25,-40){\includegraphics{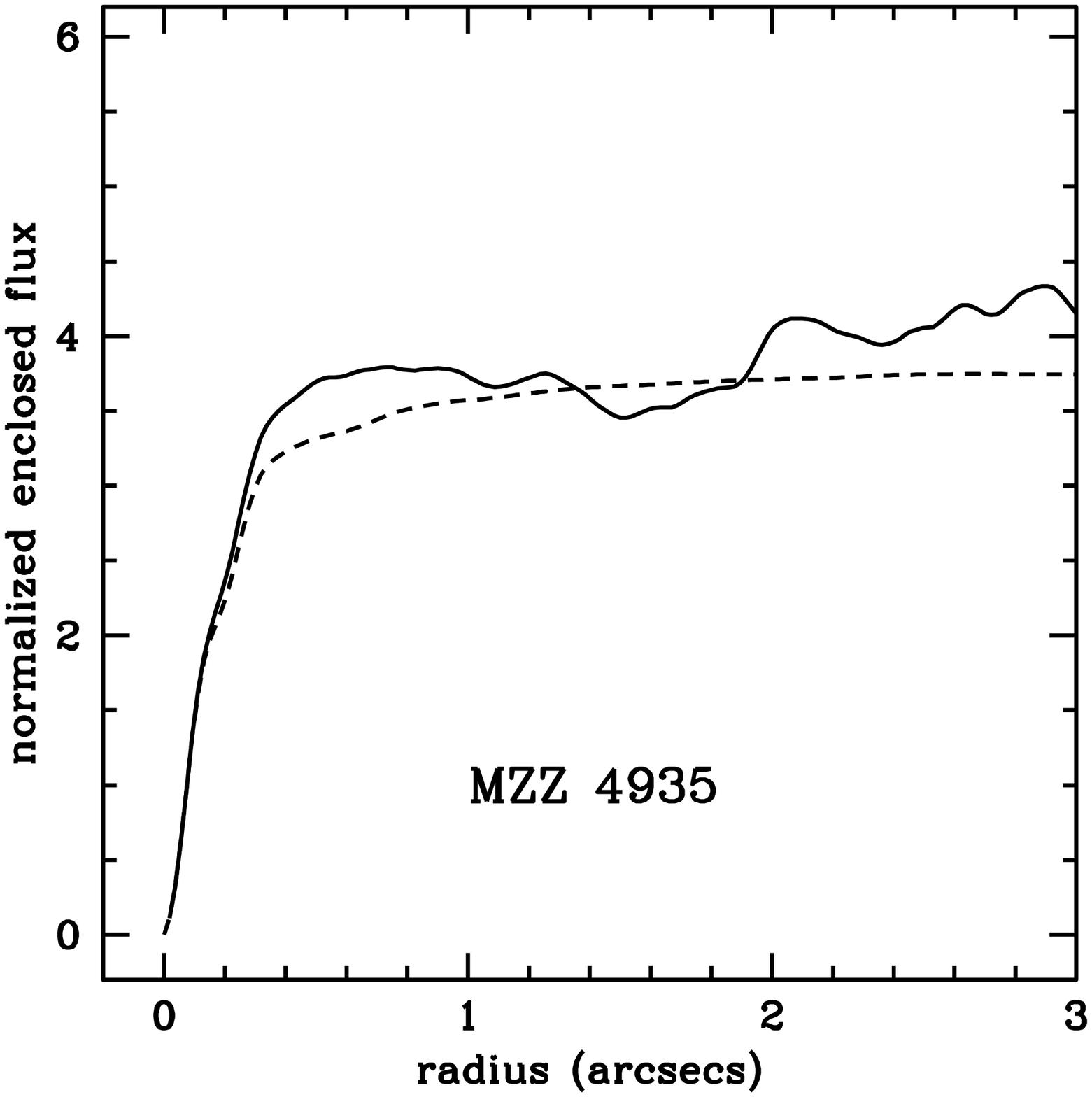}}
\put(255,-40){\includegraphics{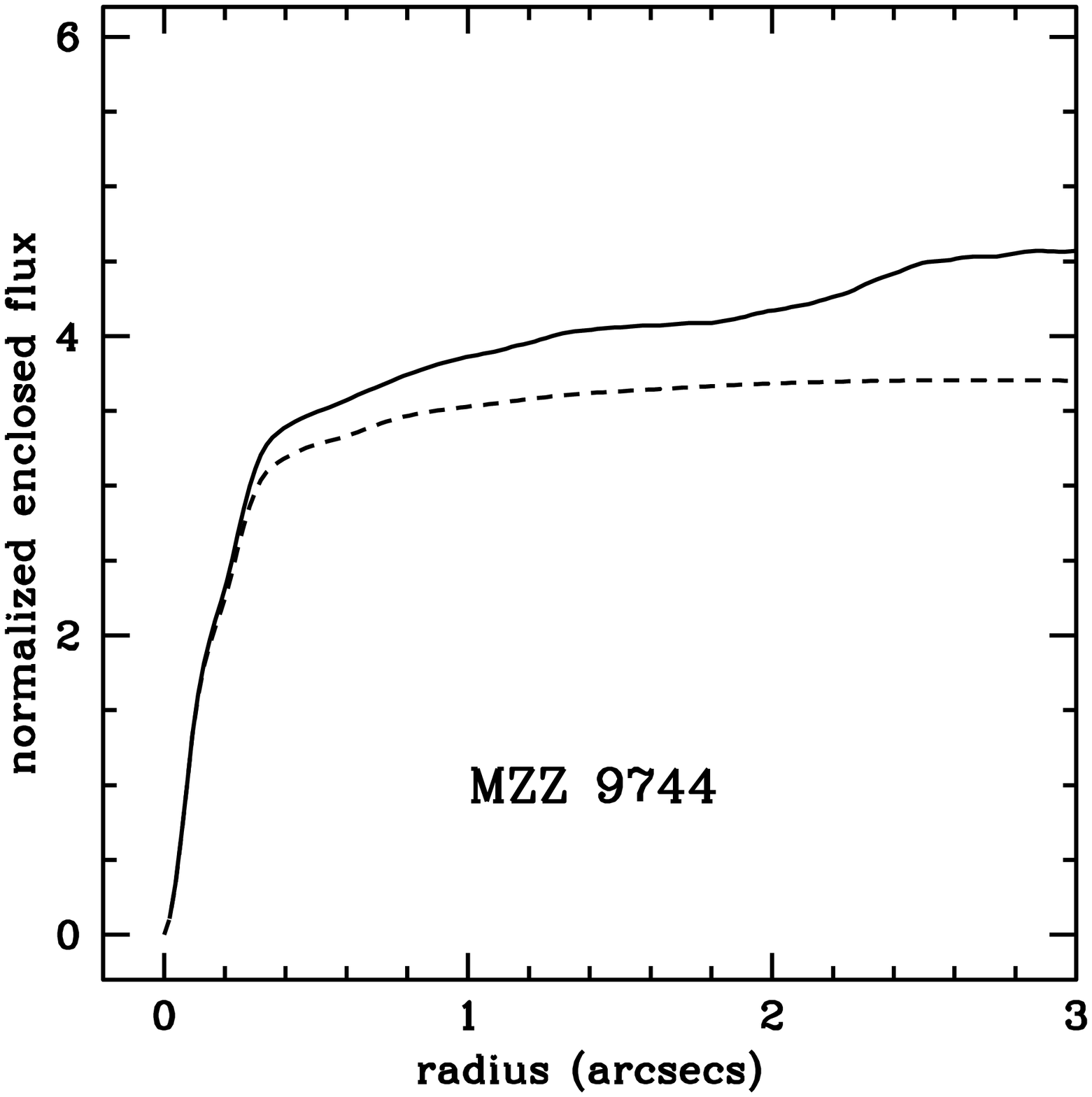}}
\end{picture}
\caption{Enclosed flux plots for 4 quasars and their PSF stars. 
The solid line is the average of the two observations of the quasar,
while the dashed line is the average of all available observations of
the corresponding PSF stars. 
The enclosed flux values are normalized to 1 at a radius of 0\farcs075 as in
Figure \ref{m9592profilefig}.
\label{profilefig}
}
\end{figure}

We display in Figure \ref{profilefig} the enclosed flux profiles for
the rest of the quasars and their corresponding PSF stars
(in these cases, averaged over the two observations). 
The plots are once again normalized such that the quasar and PSF have the same
total flux of 1 within a radius of 0\farcs075. 

We see in Figs. \ref{m9592profilefig} and \ref{profilefig}
that the 5 quasars all deviate positively from their PSF stars
(having more enclosed flux) within the inner 1\arcsec\ radius, and
that most of the excess flux is within this radius. Companion 
objects are also visible in these plots.

\begin{figure}
\begin{picture}(400,600)
\put(40,365){\includegraphics{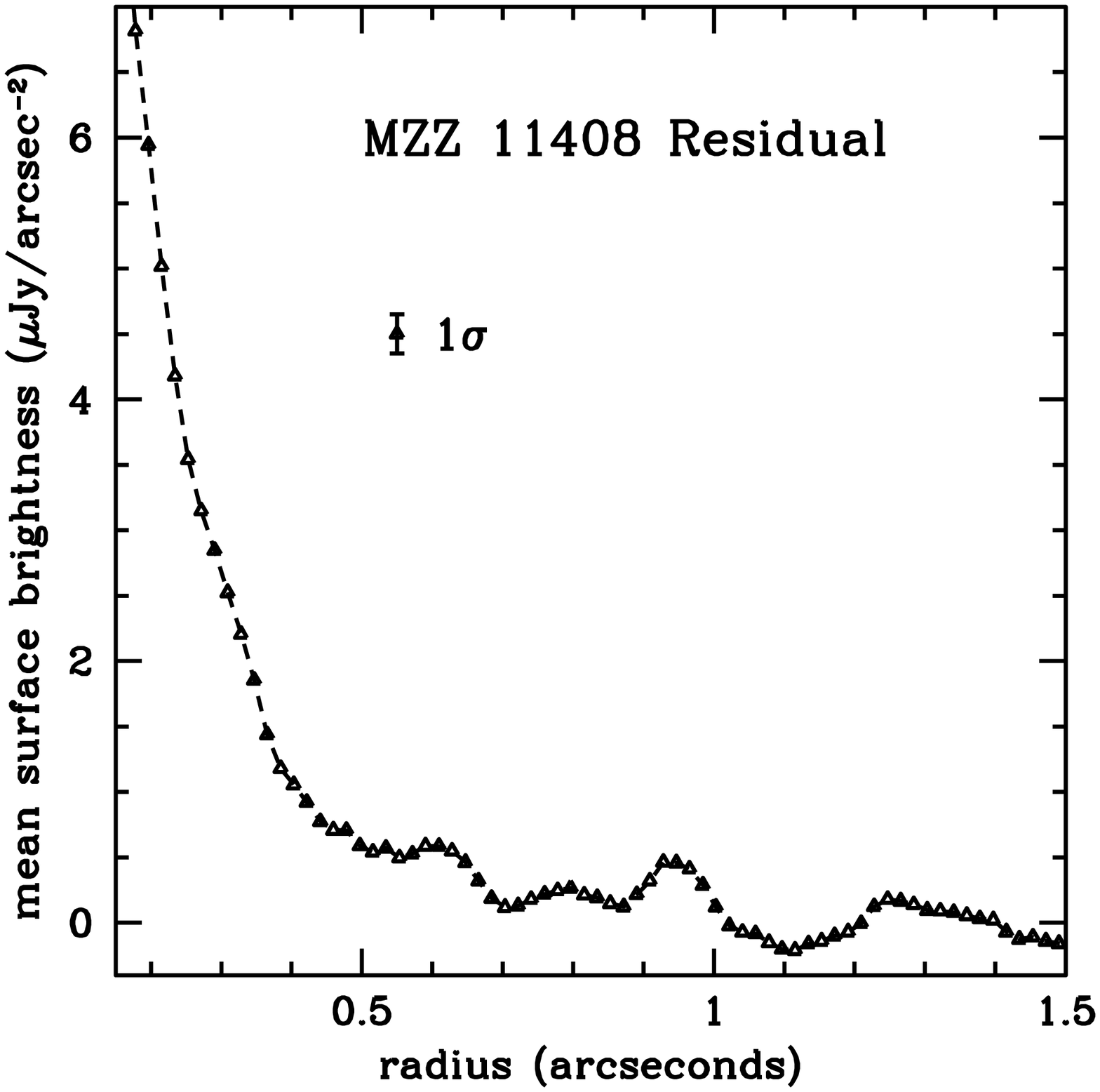}}
\put(260,365){\includegraphics{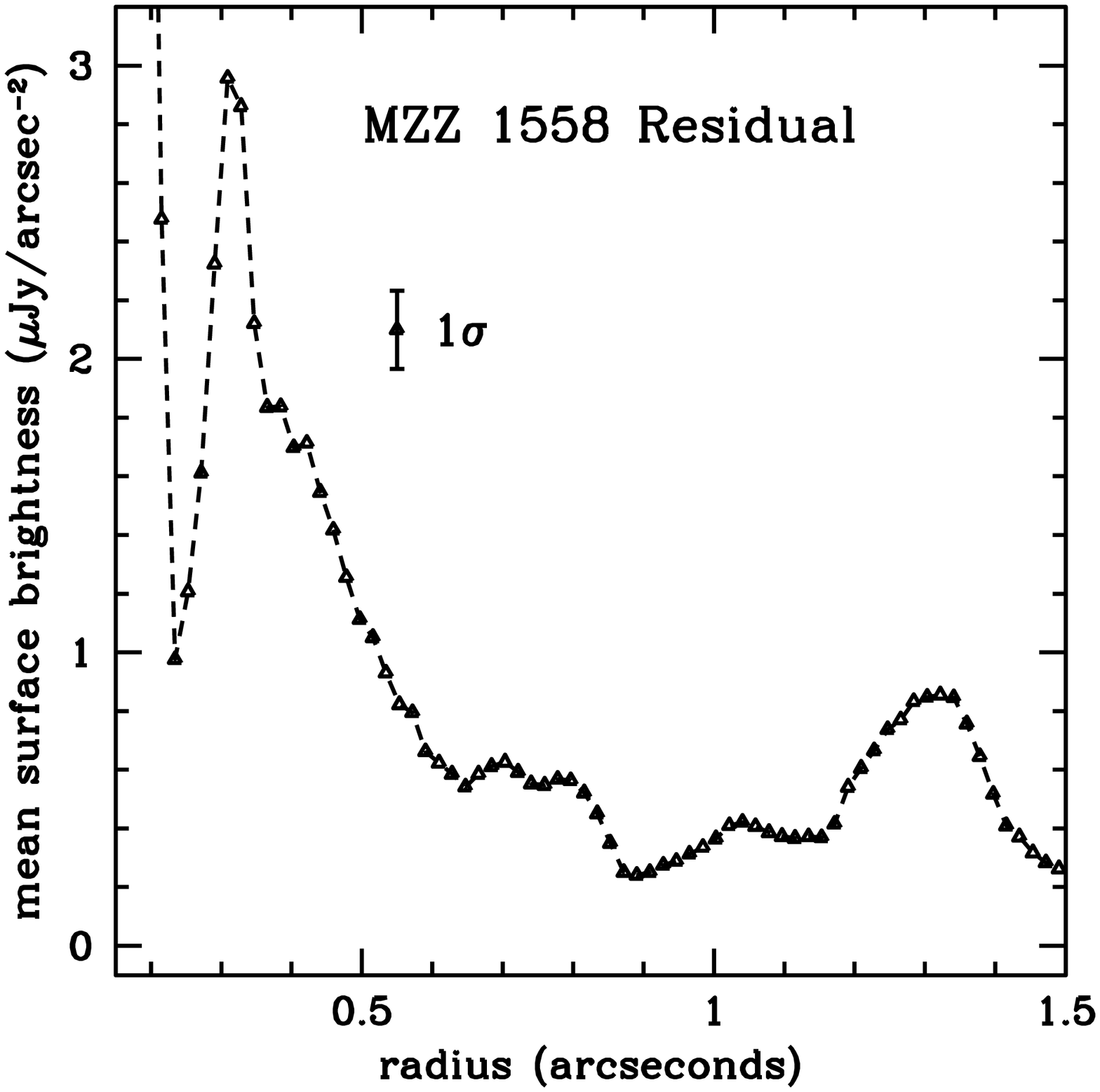}}
\put(40,160){\includegraphics{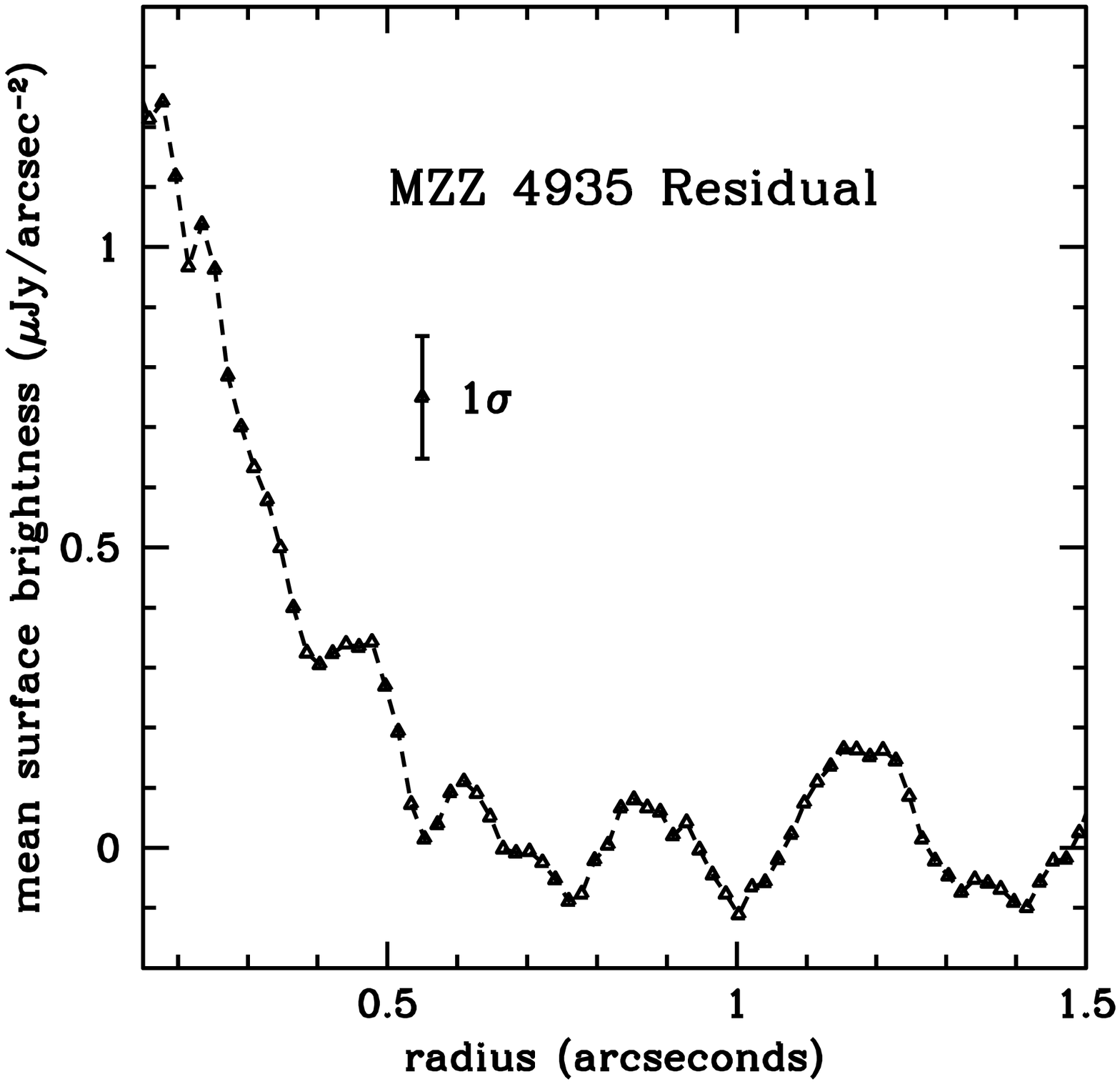}}
\put(260,160){\includegraphics{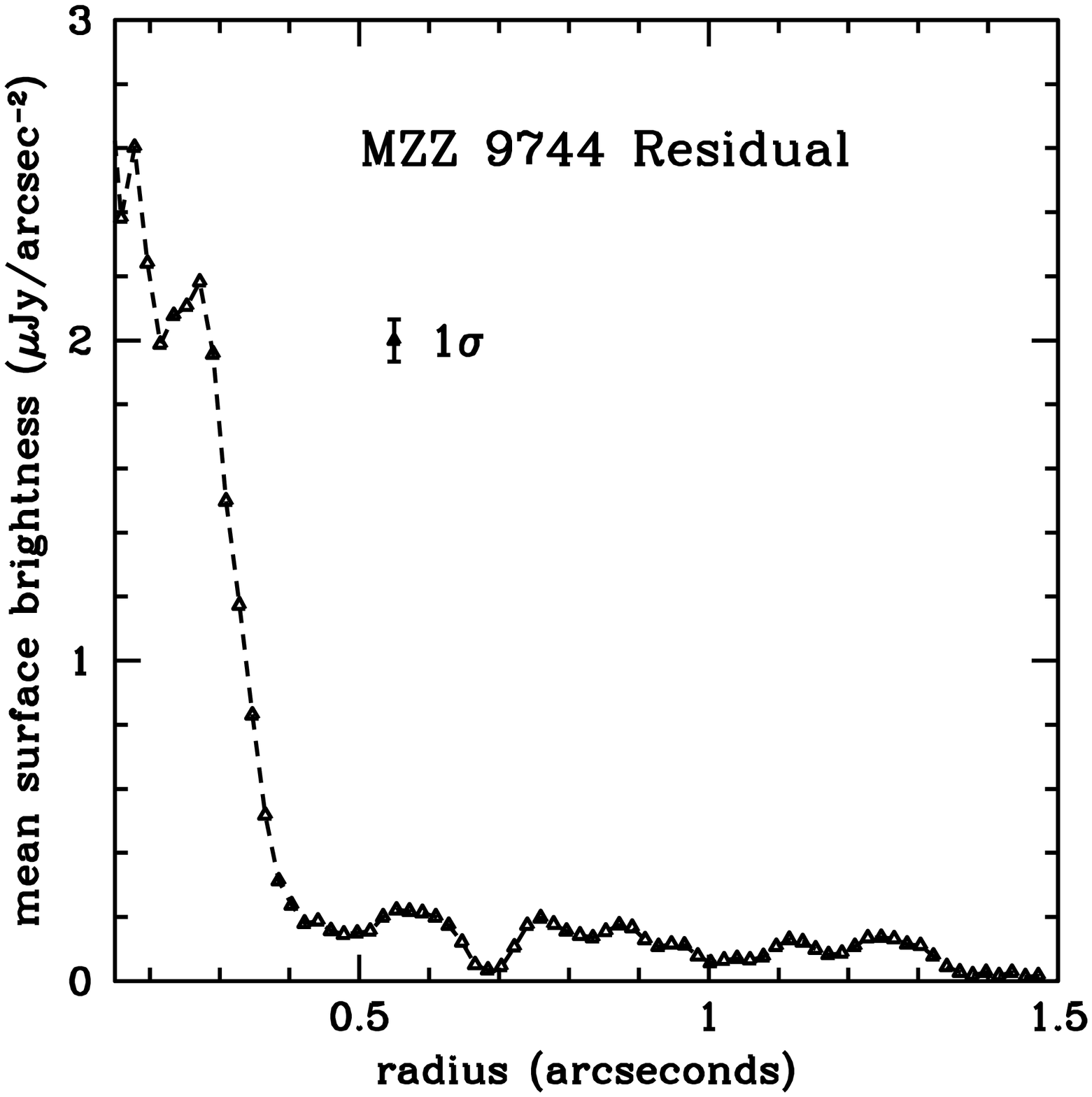}}
\put(160,-45){\includegraphics{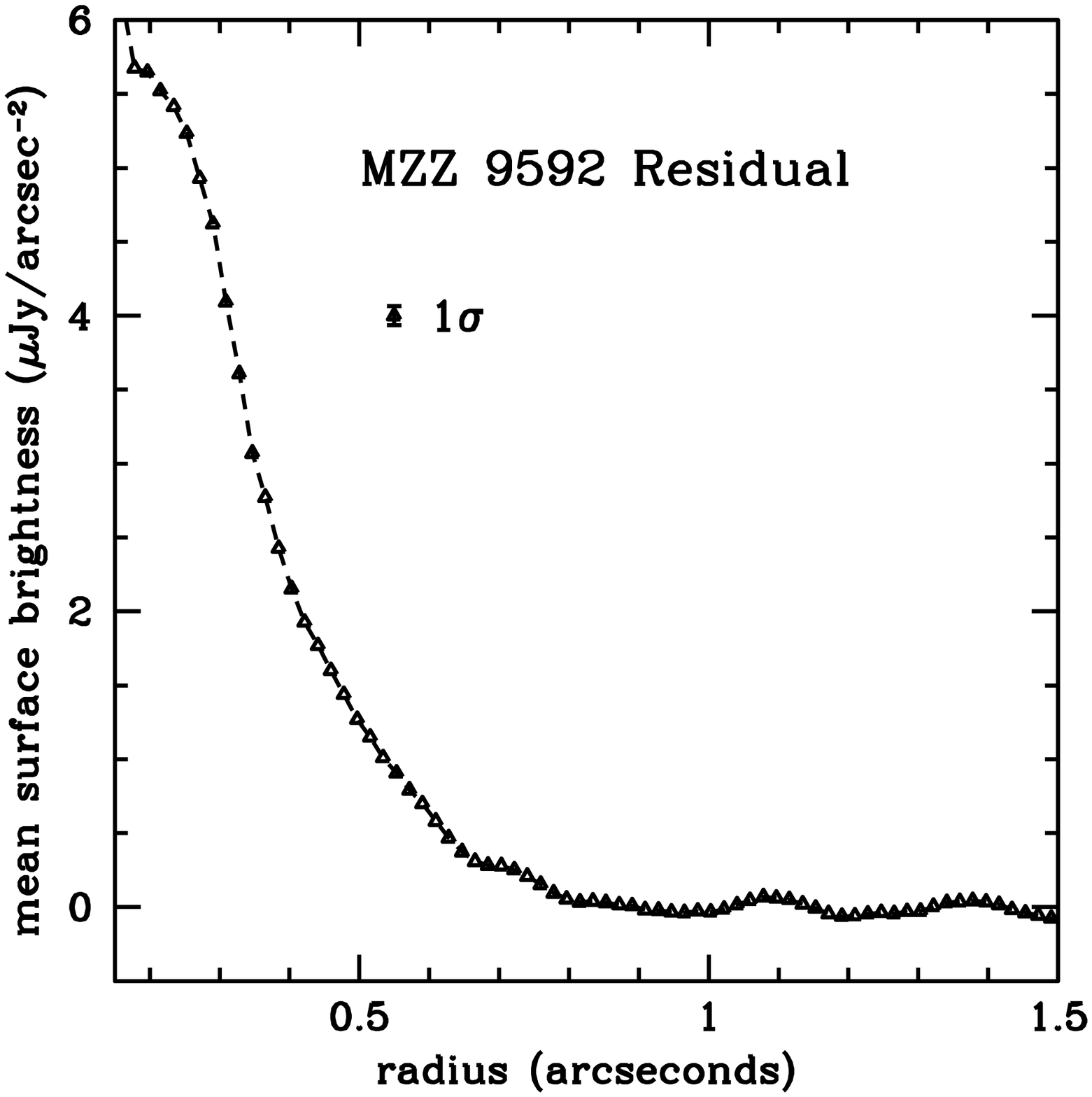}}
\end{picture}

\caption{Radial profile plots of the mean surface brightness in 
the PSF-subtracted quasar residual extensions for which the 
corresponding images are shown in Figs. \ref{z2mzzfig} and \ref{z3mzzfig}.
The radial profiles are azimuthal averages made in bins
of 0\farcs019; in each figure we show the statistical error 
due to sky noise for the average bin. 
We do not display the inner 0\farcs15 radius which
is dominated by PSF residual noise in all of the quasars.  The radial extent
of the region in which the PSF residuals dominate varies somewhat with the
brightness of the quasar nucleus.
\label{subradfig}
}
\end{figure}

To better inspect the radial behavior of the difference between the quasar and 
PSF profiles, we have also made radial profile plots of the mean surface
brightness of the final PSF-subtracted quasar residuals.  
These are shown in Figure \ref{subradfig} for the 5 quasars. 
The mean surface brightnesses are calculated in bins of 0\farcs019, and
we show in each case a statistical error for the contribution of
sky noise to the average bin. 
However, the noise between bins 
is highly correlated in these resampled, smoothed images. 

These plots show clearly the spatial extent of the azimuthally averaged
residual hosts; i.e., most flux that is not a discrete apparent
companion object is contained within the inner 1\arcsec\ radius.

\section{Results}

\subsection{Photometry}

To provide the simplest basis for cross-comparison between samples,
we have first calculated simple aperture magnitudes. We have
used apertures with radii ranging from 0\farcs22 to 1\farcs4 and
quote the values from the 0\farcs6 and 1\farcs0 radius apertures.
As these extensions
are mostly fairly compact, the smaller aperture includes most 
of the flux and reduces
sky noise, while the larger aperture, 
corresponding to an average diameter of $\sim$16 kpc at these redshifts, 
still excludes most nearby discrete companions and is more useful for
comparison to other samples of objects. 
To estimate errors, we have also calculated the host magnitudes for
the two visits separately. Though the quasar is resolved in
each of the two visits in all cases, for those objects in which
the hosts are smallest in size (e.g. MZZ 9744) or faintest (MZZ 4935)
the flux of the residual varies between the two visits
more ($\sim$40\%) than in those cases where the host
is bright and extended (e.g. MZZ 9592, visit-to-visit difference $<$5\%). 
We include these differences
in our estimates of the errors in our quoted magnitudes.

The results of our magnitude analysis for these two apertures is given in 
Table \ref{maghosttab}, and in
Table \ref{tablenuc} are the fluxes and absolute magnitudes of the quasar 
nuclei alone.  The
detected hosts vary from less than L$_*$ to 4 L$_*$, using $L_*^V$ and $L_*^B$
at $z = 0$ from the field galaxy luminosity function
of Loveday et al. (1992), i.e. $M_B^*$ = $-$21.0, and for colors of
a Sa, $M_V^*$= $-$21.8.
The errors given are generally from the systematic variation
seen between the two visits or estimated from obvious 
uncertainties in the subtraction,
although in a few cases the sky noise in the aperture was greater 
(e.g. MZZ 9592).
The faintest residual host
(around MZZ 4935) we detect at a flux $\sim$0.2L$_*$, and is 
best detected in the first visit which was less affected by persistent CR
problems.
This aperture flux corresponds to a 3$\sigma$ detection in the 
1\farcs3 diameter aperture, where the 1$\sigma$ noise was determined
by measuring the irregular background fluctuations in the sky of the 
combined frame.

These fluxes need a correction for flux lost from the
PSF subtraction process which will vary depending on how compact the
intrinsic, underlying host is. In Section \ref{sec:model}
we discuss simple models to estimate the amount of this correction.

\subsection{Morphologies}

The morphologies of these hosts are quite compact, as 
demonstrated by our aperture photometry in that most of the flux 
is contained within a 0\farcs6 radius aperture. 
A simple way of quantifying the morphologies
of these hosts is to determine the half-light radii ($r_{1 \over 2}$),
the radius at which half the flux is enclosed for each residual
host. We use as the total flux the flux enclosed within a 1\arcsec\ radius.
Inspection of the enclosed flux in the residuals as a function of
radius for each object shows a flattening of the 
enclosed energy profiles at radii of 0\farcs6 -- 1\arcsec, before contributions
from nearby companions. (MZZ 9744 has a discrete companion at 0\farcs8, and 
excluding this companion has no significant effect on the scale size derived).
These ``direct'' $r_{1 \over 2}$ values are in column (3) of 
Table \ref{proffittab}.
By making these estimates of the half-light radii we have essentially 
assumed that the flux profile within the inner 0\farcs2 radius is flat, 
and these values may therefore be overestimates if the flux profile is more
sharply peaked. 
Statistical error in the enclosed flux results in only a 
small statistical error in $r_{1 \over 2}$, $\sim$0\farcs02, but systematic
errors caused by assumptions about the profile and errors in
PSF subtraction may be greater.  

In addition, we make fits to the azimuthally-averaged
radial profiles of the two brightest residual hosts, using simple 1D exponential
disk and de Vaucouleurs profile models to extrapolate the central 
0\farcs2 radial region. 
The half-light radii resulting from these
fits are in column (4) of Table \ref{proffittab} for
MZZ 9592 and MZZ 11408,
in which the residual hosts are least affected by the PSF subtraction
process. 
(For the de Vaucouleurs profile, $r_{1 \over 2}$ is equal to the 
effective radius $r_e$, while for the exponential disk model, 
$r_{1 \over 2}$ is 1.6783 times the exponential scale length.)
The fits were made in the radial region $0\farcs23 < r < 1\farcs2 $
to exclude the inner PSF-dominated disk, and reduce sky noise.
The range in scale sizes quoted are a result of
varying slightly the
fitting radial regions while still maintaining a reasonable fit.
These fits are intended as a cross-check on the validity of the 
direct $r_{1 \over 2}$ derivations, and 
the results are consistent.

The scale sizes (from all methods) are securely less than
$\sim$0\farcs5 ($\sim$4 kpc), and the mean scale size based on the
direct $r_{1 \over 2}$ determinations is $\sim$2.3 kpc in physical scale.

\begin{figure}[!hp]
\begin{center}
\begin{picture}(400,500)
\put(-100,-40){\includegraphics{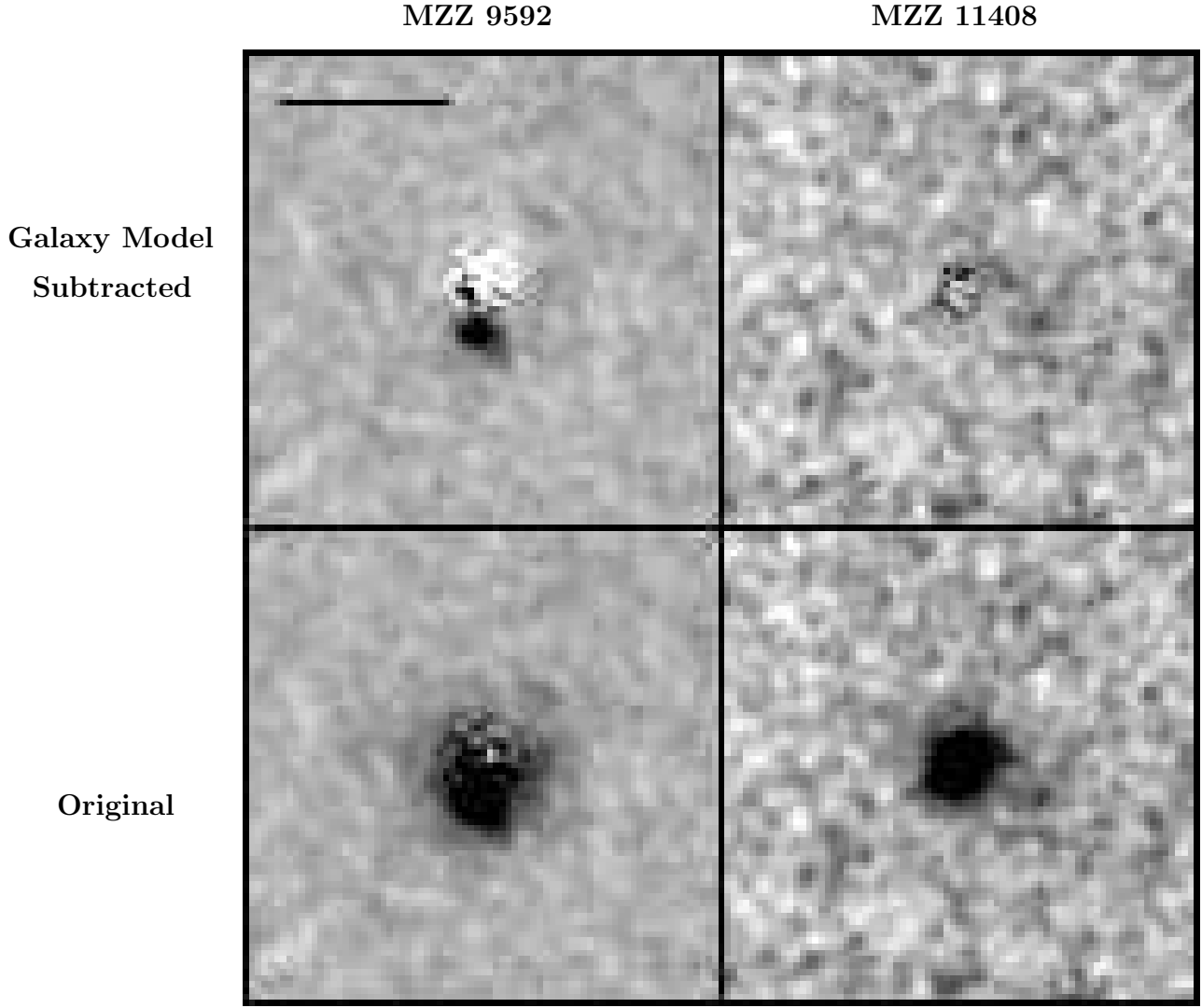}}
\put(-40,150){\large{\bf Original}}
\put(-60,380){\large{\bf Galaxy Model}}
\put(-50,360){\large{\bf Subtracted}}
\put(100,470){\large{\bf MZZ 9592}}
\put(290,470){\large{\bf MZZ 11408}}
\end{picture}
\caption{Unsmoothed images of
the PSF-subtracted quasars MZZ 9592, left, and MZZ 11408, right. The 
bottom panels are the PSF-subtracted residual host images, while in the 
upper panels, a two-dimensional model of a galaxy has been 
subtracted as well. 
For MZZ 9592, this was a disk model with a $r_{1 \over 2}$
of 0\farcs25; for MZZ 11408, this was a bulge model with $r_{1 \over 2}$
of 0\farcs3. 
The black bar shows 1\arcsec.
\label{modelsub}
}
\end{center}
\end{figure}

In two cases (MZZ 1558 and MZZ 9744), there are galaxies close to the quasar in projection ($\lesssim$
10 kpc), and in one (MZZ 9592) there is an off-center component in the residual host.
In MZZ 1558, at $z = 1.829$, the companion galaxy is at a 
projected distance of 1\farcs3 (11 kpc) from the quasar, and 
has a 1\farcs3 aperture
flux of 2.2 $\mu$Jy, while the quasar host itself has a flux of 10\% less.
This could therefore be an example of an early stage of an equal mass merger. 
In the case of MZZ 9744, the nearest companion in projection is
of much less flux than the host.
Assuming these companion galaxies and residual component
are associated, some of these 
hosts may not be completely relaxed systems, and could be
in some stage of a merger.
However, other than these fairly discrete apparent companions and residual component,
the hosts appear regular.
After subtraction of a regular 2D galaxy model from the two brightest hosts,
the off-center component in MZZ 9592 is more visible and appears compact;
in contrast, a similar subtraction from MZZ 11408 leaves
no irregular components (Fig. \ref{modelsub}). 

\section{Comparison to other samples of galaxies: simulations}

\label{sec:model}

In order to compare this sample of galaxies to other samples of observed
high-$z$ galaxies, we need to understand how the presence of
the quasar nucleus and the PSF-subtraction procedures as well
as the observational parameters (such as exposure time and resultant
sky noise) affect our ability to measure the magnitudes and morphologies
of the hosts. The simplest way of addressing this issue is to simulate
our quasars with other observed galaxies, as well as noiseless models.

\subsection{Quasar simulations with NICMOS-observed HDF-N Lyman break galaxies}

\label{sec:lbgsim}
We have made some simple simulations
of the effect the presence of the quasar nucleus and
the PSF-subtraction process have on the
derived magnitudes and morphologies. 
We have used for the first set of these
simulations a sample of spectroscopically-identified
$z \sim$ 2 -- 3 Lyman break galaxies (LBGs) from the HDF North
(Dickinson 2000). We were kindly provided with the drizzled, calibrated
NICMOS image by Dickinson et al. (2000). This NIC3 F160W image
was observed to a much greater depth than our observations.

We made ``artificial quasars'' at two fiducial redshifts, $z = 1.8$ 
and $z = 2.7$, to simulate our observed data. 
To do this, we used cutout images of the Lyman break galaxies in this sample
that are within about 0.3 redshift units of these two redshifts,
and correct their fluxes and scale sizes for 
cosmological effects, giving us a sample of 12 objects for the 
$z = 2.7$ models, and 6 objects for the $z = 1.8$ models. 
After resampling these imaging data to match our pixel scale, we
added to the galaxy center
nuclei of varying magnitudes (matching the range of
magnitudes observed),
using the MZZ 9592 field star as the nucleus. 
(We also ran a set of simulations 
using a selection of observed PSF stars as the nuclei).
Background and Poissonian noise were added to match the average noise
seen in the observed quasar images, with the limiting surface brightnesses
normalized to the model redshifts of $z = 1.8$ and $z = 2.7$.

\begin{figure}[p]
\begin{picture}(400,300)

\put(-190,-442){\includegraphics{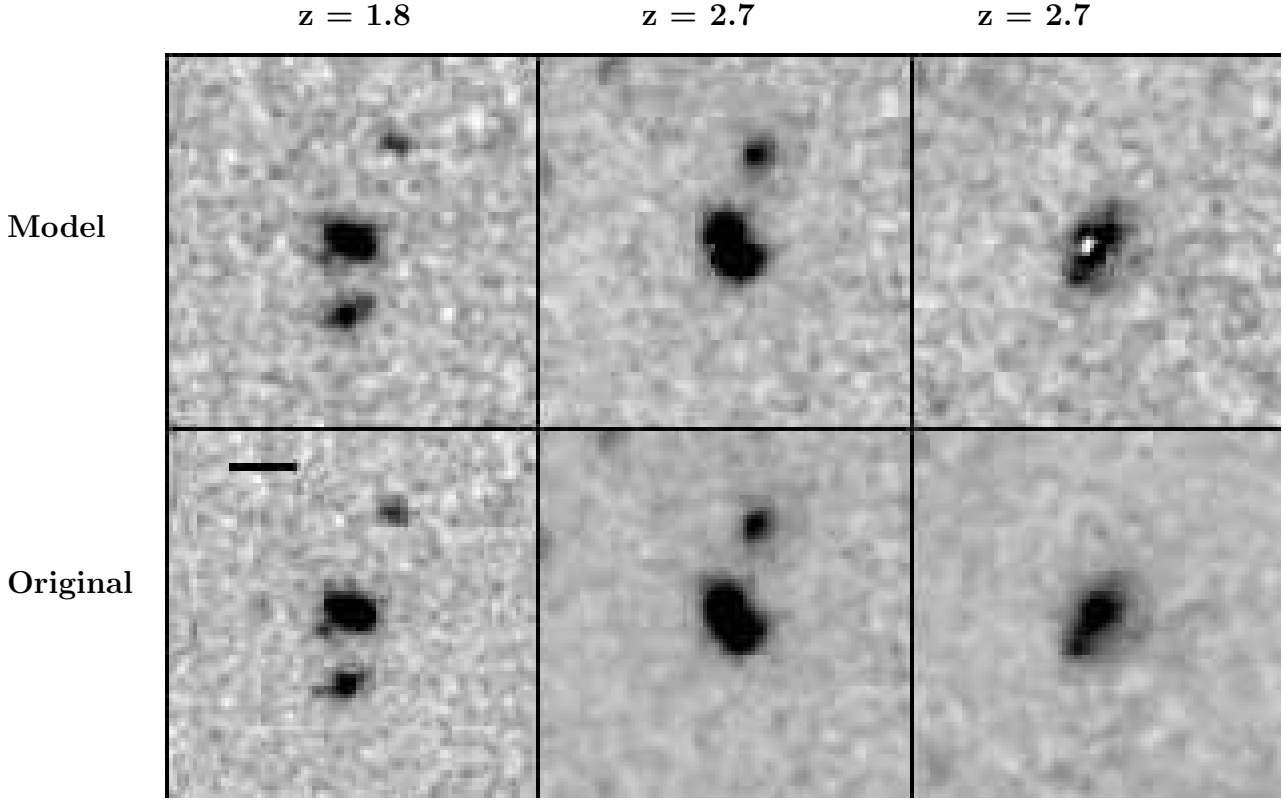}}
\put(0,90){\large{\bf Original}}
\put(0,225){\large{\bf Model}}
\put(110,305){\large{\bf z = 1.8}}
\put(240,305){\large{\bf z = 2.7}}
\put(367,305){\large{\bf z = 2.7}}

\end{picture}

\caption{Examples of simulated quasars based on NICMOS-observed
Lyman break galaxies: 3 HDF North galaxies from Dickinson et al. (2000), 
used for one 
$z = 1.8$ model and two $z = 2.7 $ models. 
The bottom row shows the original galaxy before nucleus and noise is added,
while the top row shows the result of adding
artificial quasar nuclei, sufficient Poissonian 
noise to match our observed quasars, and then making a standard subtraction of
an independent observed PSF. 
The artificial nuclei used for the models, shown from left
to right, were of low, medium and high brightnesses respectively, 
as given in Table \ref{modelmagtab}.
The black bar represents 1\arcsec, and each frame
is 5\farcs7. The display 
stretch is the same as that used in Figures \ref{z2mzzfig}
and \ref{z3mzzfig}. 
\label{hdfmodelfig} }
\end{figure}

\begin{figure}[p]
\plottwo{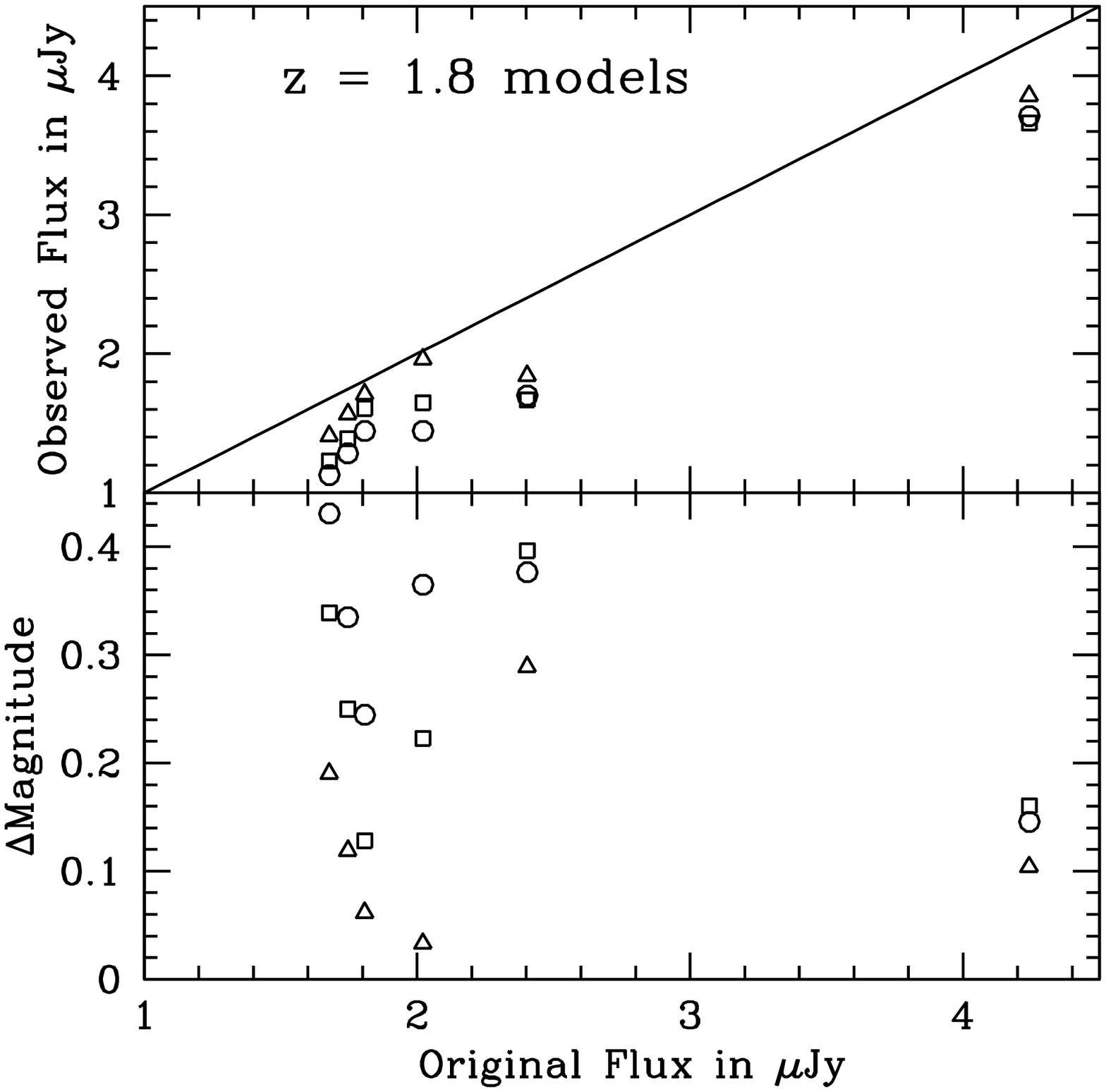}{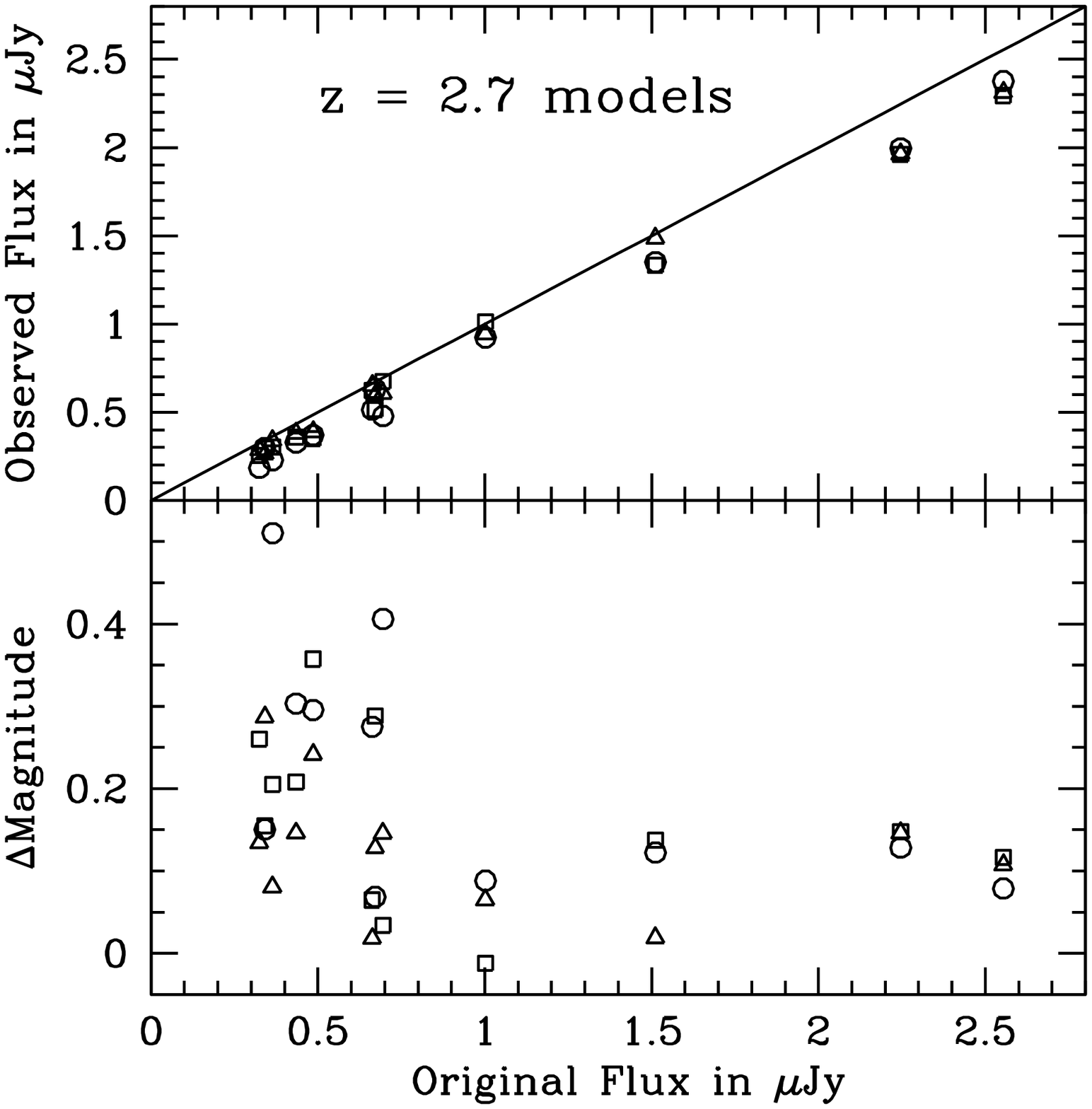}
\caption{
Top panels, 
left and right, are plots of output flux in a 1\farcs3 aperture after 
application of our modelling and PSF-subtraction method versus the original
flux in the input galaxy in the same aperture. Bottom, we give the corresponding
difference in host magnitude. Each input galaxy is 
modelled with nuclei of three different brightnesses: the circle
shows the flux from using the 
brightest nucleus, the square corresponds to
the medium nucleus, and the triangle shows the result from
using the faintest nucleus. (These model nuclear brightnesses are given
in Table \ref{modelmagtab}).
\label{modellinefig} }
\end{figure}

After running a standard PSF-subtraction using 
the PSF star observed in the same visit as the field
star, we have calculated the
flux in the same apertures used for the real quasars,
both in the PSF-subtracted simulated quasar and the original galaxy used for each model. 
In Figure \ref{hdfmodelfig}, we show 
a few examples of HDF Lyman break galaxies before and after
the modelling process. In Figure \ref{modellinefig} we show the results 
of the photometry of the $z = 1.8$ and $z = 2.7$ model samples.
The success of the process in recovering the input
flux of the galaxy varies with the flux and scale size
of the original
galaxy, and with the effective brightness of the overlying quasar.
We give the input parameters of the models and the resultant
$\Delta$ magnitudes in Table \ref{modelmagtab}, where we also indicate which model
is simulating which of our 5 quasars. 
From inspection of
these simple models, we find
first, that the bulk of the flux offset occurs in the central
0\farcs5 diameter section; second, that the addition of 
the noise and the specifics of the PSF -- nucleus profile 
mismatches can cause small offsets in the flux
either way, although 
the trend for most models with peaked galaxies is for the 
subtraction process to decrease the flux. 
In addition, the brighter the overlying nucleus, the larger the
decrease in the derived flux.

In the plots shown
here, we used the MZZ field star as the ``nucleus'' of the model quasar and
the observed PSF star from the same visit as the subtraction PSF, which
should be the most representative of an observed quasar nucleus/PSF star
pair. Using the
various PSFs as the nuclei and subtracting other PSFs from different visits
gave results with similar trends, but with more scatter,
as the PSF residuals are significantly larger if they are not observed in the
same orbit. 

\subsection{Quasar simulations with observed radio galaxies}

We have also made equivalent models with some NICMOS F160W and F165M
images of powerful radio galaxies at $z$ = 2 -- 3 (Pentericci et al.
2000; Pentericci 2000) 
selected primarily from the MRC sample of McCarthy et al. (1996).  
Taking from their sample only the observations of galaxies that are 
completely emission line-free and resolved would limit us to only two objects. 
We have also therefore
included galaxies that have relaxed elliptical morphologies;
these galaxies are less likely to have major contributions from
extended emission-line regions. 
The final sample comprises 4 $z \sim 2$ and 3 $z \sim 3$
radio galaxies. 

\begin{figure}[p]
\begin{picture}(400,300)
\put(-190,-442){\includegraphics{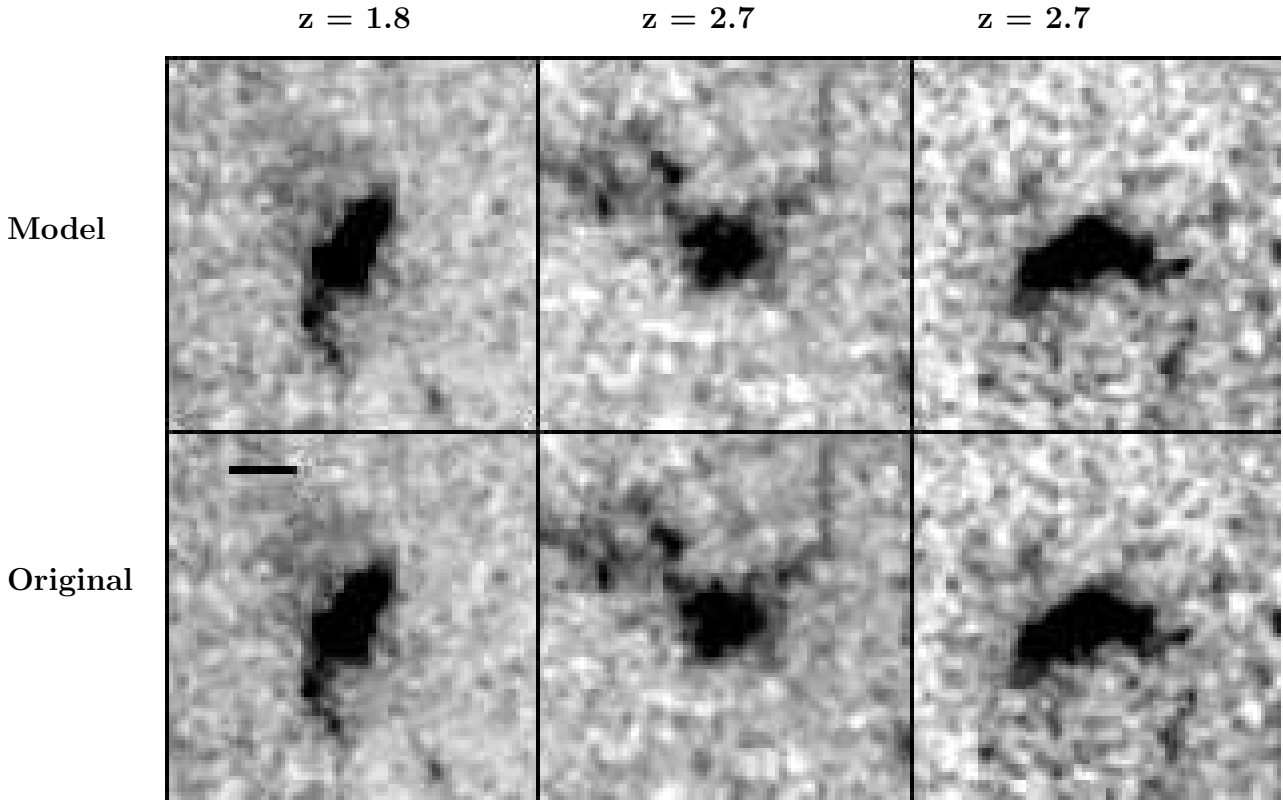}}

\put(0,92){\large{\bf Original}}
\put(0,225){\large{\bf Model}}
\put(110,305){\large{\bf z = 1.8}}
\put(240,305){\large{\bf z = 2.7}}
\put(367,305){\large{\bf z = 2.7}}
\end{picture}

\caption{Example simulated quasars based on NICMOS-observed powerful
radio galaxies: 3 MRC galaxies from Pentericci et al. 2000, used for one
$z = 1.8$ model and two $z = 2.7 $ models.
The bottom row shows the original galaxy before nucleus and noise is added,
while the top row shows the result of adding
artificial quasar nuclei, sufficient Poissonian
noise to match our observed quasars, and then making a standard subtraction of
an independent observed PSF.
The artificial nuclei used for the models, shown from left
to right, were of low, medium and high brightnesses respectively, 
and are listed in Table \ref{modelmagtab}.
The black bar represents 1\arcsec, and each frame
is 5\farcs7. The display stretch for the $z = 2.7 $ images
is the same as that used in Figures \ref{z2mzzfig}
and \ref{z3mzzfig}, while the $z = 1.8$ model images (at left) are shown with
10$\times$ less stretch. 
\label{rgmodfig} }
\end{figure}

We apply to this sample the same modelling process described in \ref{sec:lbgsim},
A few examples of the results are shown in Fig. \ref{rgmodfig};
in essence, this is how a few typical powerful radio galaxies at
$z \sim 1.8$ and $z \sim 2.7$ would appear had they contained
quasar nuclei similar to those of our quasars and had been observed 
to almost the same depth. 
In some of these radio galaxies (as discussed in Pentericci et al.
2000) there is a large intrinsic contribution from 
unresolved nuclei; this 
unresolved nuclear component is removed by the subtraction
process and results in a large change in magnitude. 
As can be seen from Table \ref{modelmagtab}, the $\Delta$ magnitudes 
are larger than for the Lyman break galaxies, $\sim$0.3 and 0.4 mags for 
the $z = 1.8$ and $z = 2.7$ models. 

\subsection{Comparison to low-$z$ quasars: bulges and disks}

\label{sec:lowzsim}
In order to make a good comparison of the 
high-$z$ quasar hosts to the well-studied low-$z$ quasar hosts,
we have made another set of simulations. In this case, we wished to
estimate how a typical low-$z$ quasar host would appear
at $z = 1.8$ and $z=2.7$, with nuclei of comparable brightnesses to ours and
imaged under similar observational conditions.
From these simulations we can also determine whether our seemingly compact, 
moderate-luminosity hosts
could simply be the tips of typical, larger scale-size elliptical or disk
galaxies, in which most of each galaxy has been obscured by sky noise.

Bahcall et al. (1997) and McLure et al. (1999) both
have imaged samples of low-$z$ ($z \sim 0.2$) quasars 
with WFPC2 at $\sim R$-band, giving absolute magnitudes and scale sizes of
the hosts at close to rest-frame $V$, similar to the rest-frame
of our high-$z$ $H$ band observations.
We therefore convert Bahcall et al.'s total absolute $V$ magnitudes (from
their Table 12) to our cosmology, and their scale sizes for fitted disk and
bulge models to half-light radii, and find that their average 
(of 14) radio-quiet quasar hosts has an absolute $V$ magnitude 
$M_V$=$-$22.1, with a 1$\sigma$ variation of 0.6. The average 
$r_{1 \over 2}$ is 8.1 kpc. 
We have also used the total $R$ magnitudes
for the radio-quiet quasar sample of McLure et al. (1999)
(from their Table 2).
We converted these to $M_V$ and found the average RQQ host to have
$M_V$=$-$22.7 with 0.6 mag variation,
and half-light radius of 8.2 kpc (if one host with
$r_{1 \over 2}$ = 23 kpc is excluded). For our models, therefore, 
we have adopted a $r_{1 \over 2}$=8.2 kpc as typical, and have
used both $M_V$=$-$22.1 and $-$22.7 for a reasonable spread
in brightnesses.

\begin{figure}[!hp]
\begin{picture}(400,300)
\put(-190,-447){\includegraphics{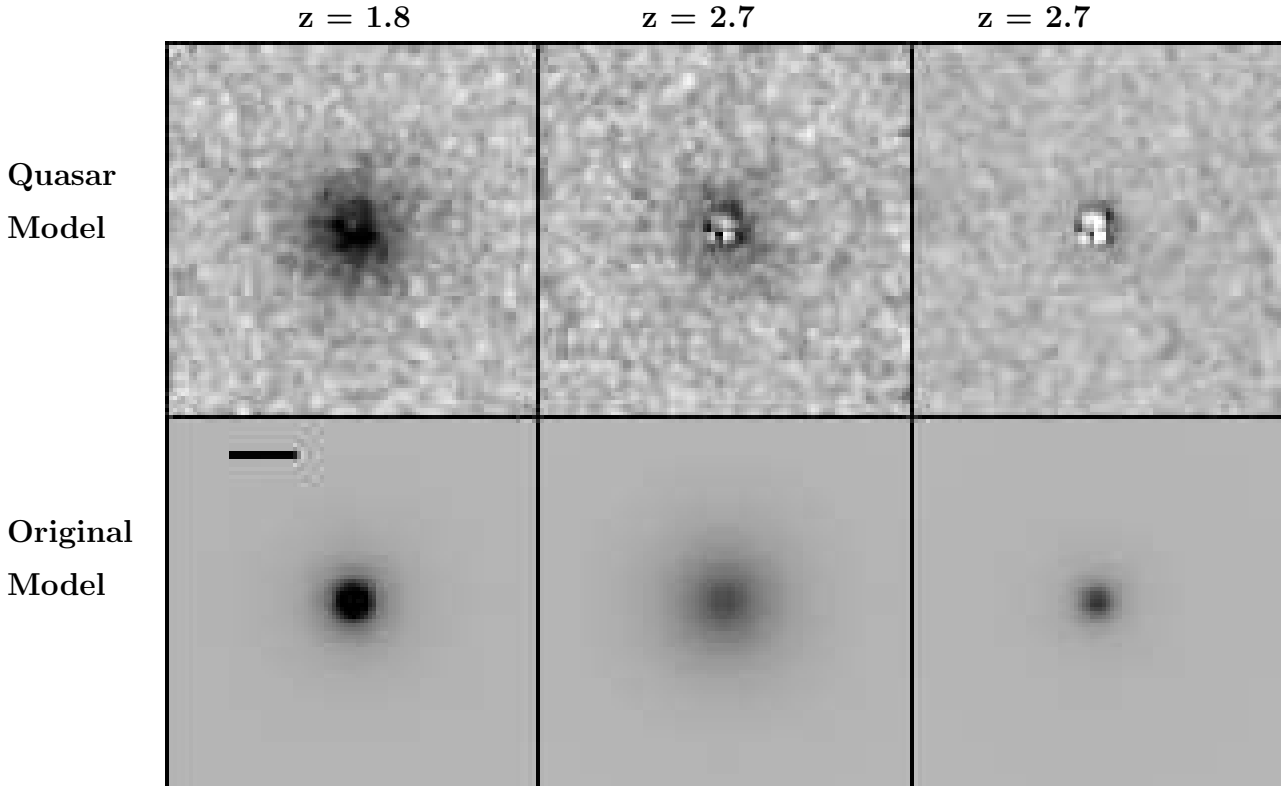}}
\put(0,100){\large{\bf Original}}
\put(0,80){\large{\bf Model}}
\put(0,235){\large{\bf Quasar}}
\put(0,215){\large{\bf Model}}
\put(110,295){\large{\bf z = 1.8}}
\put(240,295){\large{\bf z = 2.7}}
\put(367,295){\large{\bf z = 2.7}}

\end{picture}

\caption{Example quasars simulated from disk and 
elliptical models, with half-light radii of 8.2 kpc and $M_V$ = $-$22.1: 
two $z = 1.8$ models and one $z = 2.7 $ model.
The bottom row shows the original galaxy model before nucleus and noise is added,
while the top row shows the result of adding
artificial quasar nuclei, sufficient Poissonian
noise to match our observed quasars, and then making a standard subtraction of
an independent observed PSF.
The artificial nuclei used for the models, shown from left
to right, were of low, medium and high brightnesses respectively. 
These values are given in Table \ref{modelmagtab}.
The galaxy model types used were elliptical, disk, and elliptical models 
from left to right. 
The black bar represents 1\arcsec, and each frame
is 5\farcs7. The display
stretch is the same as that used in Figures \ref{z2mzzfig}
and \ref{z3mzzfig}.
\label{ellmodfig} }
\end{figure}

We generated noiseless galaxy models with both $r^{1 \over 4}$-law 
and exponential disk radial profiles with half-light radii
of 8.2 kpc, which corresponds to 0\farcs98 and 0\farcs92 at
$z = 1.8$ and $z = 2.7$ respectively. 
After scaling the model hosts to the appropriate apparent 
magnitudes and making the slight necessary $K$-corrections,
we convolved the model galaxies with the seeing disk of our observed
PSFs, and simulated the observed quasars in the same
way described in the previous sections. 
In Figure \ref{ellmodfig} are shown a few examples of the resultant
images, all for the fainter absolute host magnitude ($M_V$ = $-$22.1).
In terms of the amount of flux recovered, 
the disk model galaxies fare slightly better
than the ellipticals, but the difference is minor. The
$\Delta$ magnitudes given in Table \ref{modelmagtab} are therefore
averaged over the model types and brightnesses.
The total host flux lost is not significant ($\sim$ 0.1 -- 0.2
mags) at $z = 1.8$, though is worse 
for the brighter nuclei at $z = 2.7$ ($\sim$0.3 -- 0.4 mags). 

To estimate how much our scale size determinations could be affected
by the PSF-subtraction and and sky noise, we have also made 1D azimuthal
averages of the model galaxies before and after the modelling process,
and attempted to fit these residual profiles directly.
We find that the
fitting is not always stable,
but most of the fitting to the residuals
reflects accurately the relatively large scale sizes of the input models. 
In the worst case, however, of the fainter elliptical host at $z = 2.7$ with the
brightest of the model nuclei, we did find incorrect, compact
scale sizes. This is the model shown to the far right of Figure 
\ref{ellmodfig} in which 1D residual fits found a half-light radius of $\sim$0\farcs2.
In this case, too much of the extended material has been lost to sky noise to make
an accurate determination of the original scale size from the residual.
However, except 
for the highest redshift objects with the brightest nuclei, these models indicate
that we can probably differentiate the compact MZZ hosts easily from a noisy
host galaxy with larger scale size. 
The smoothness of the input host model accentuates this problem, and
determinations of scale sizes in the Lyman break simulations were
more successful. 

\subsection{Aperture and subtraction corrections}

\label{sec:apert}

To estimate the total magnitudes in our hosts we make two corrections
to our aperture magnitudes: a ``subtraction'' correction, to estimate the amount
of flux lost to the PSF subtraction process, and a simple aperture correction,
for the amount of flux that will lie outside our aperture.

The subtraction correction is the greater source of uncertainty, and will
depend strongly on how compact the inner regions of the underlying galaxy are,
and on the brightness of the overlying quasar nucleus. 
We have summarized the decrements in host magnitude that resulted from 
our various simulations in Table \ref{modelmagtab}, as a function
of galaxy type, galaxy redshift, and quasar nuclear magnitude.
The Lyman-break galaxies are compact and faint, and of the three types
of galaxies (LBGs, radio galaxies, and low-$z$ quasar hosts) modelled,
are the most comparable to the MZZ hosts. 
We therefore use as the subtraction correction for each host
the average $\Delta$ magnitude found from the LBG models.
The quasar name and the corresponding $\Delta$ magnitude used are found in
columns (4) and (5) respectively of Table \ref{modelmagtab}.

For the aperture corrections, we have assumed that our host galaxies
are pure disks or ellipticals with the scale sizes we have determined
from the residuals, and have estimated the amount of flux that lies
outside of our standard 2\arcsec\ apertures.
The scale sizes were primarily in the range of 0\farcs2 -- 0\farcs4.
The corresponding range in magnitude corrections for the elliptical models are 
0.14 -- 0.39 mags for the 2\arcsec\ aperture. For the disk
models, the magnitude corrections necessary are smaller, with a range of
0.0 -- 0.18 mags for the 2\arcsec\ aperture. 
We quote in column (5) of Table \ref{proffittab} the aperture corrections
we adopted for each quasar. 

From a combination of these subtraction and aperture effects, we estimate
a probable correction from the 2\arcsec\ aperture to a total magnitude for
each host, and give this corrected ``total'' magnitude
in the last column of Table \ref{maghosttab}.

\section{Discussion}

We have detected host galaxies around 
all five of our sample of $z \sim 2$ -- 3 radio-quiet quasars. 
The detected hosts have absolute $V$ magnitude $M_V$ $\sim$ $-$20.2 to
$-$22.5, with absolute luminosities from less than present-day $L_*$ to 4 $L_*$.
The quasar nuclei are of relatively faint absolute $B$ magnitudes, with $M_B$
ranging from $-$24 to $-$22.  
The scale sizes, where we can fit the residuals, 
yield half-light radii of $<$4 kpc.

\subsection{Host galaxy magnitudes compared}

Observational resolution is obviously
important in studying the hosts of quasars, particularly
those at high redshift, since
(at $z$ = 2) $\sim8$ kpc of physical scale
would be hidden under a typical $1\arcsec$-diameter groundbased seeing disk.
We will thus compare our work primarily to the results of other HST imaging;
however, work has been done on high-$z$ quasars with ground-based
resolution that provides tantalizing hints of possible correlations.
While Lehnert et al. (1992) found bright hosts around 6 $z \sim 2.5$
radio-loud quasars in ground-based near IR imaging, Lowenthal et al. (1995)
detected no hosts
around a comparison sample of RQQ at the same $z$ and nuclear magnitudes. 
This result implied that the host luminosity might correlate with
radio-loudness.  In contrast, Aretxaga et al.\ (1995, 1998) 
studied several very luminous $z \sim 2$ RQQs and found hosts
comparable in luminosity to those found in the RLQ sample of Lehnert et al.
(1992), indicating perhaps a heavy dependence of host magnitude on the
nuclear luminosity. 

Our observations of the $z \sim 2$ -- 3 RQQ sample benefit from
the compact and stable NICMOS PSF.
Indeed, our simple simulations with observations of other high-$z$ galaxies
have indicated that for most intrinsic galaxy morphologies 
our estimates of the magnitudes and morphologies
are not badly affected by the presence of the quasar nucleus and observing
process. 

We have quoted in Table \ref{maghosttab} simple aperture magnitudes,
with error estimates based on the variation in the subtraction and 
from the sky noise. In the last column, we also give a rough estimate of 
``total'' magnitude, 
based on adding the systematic subtraction and aperture corrections discussed
in section \ref{sec:apert} to the 2\arcsec\ aperture magnitudes. 
It is these estimates of total magnitude that we will use in comparing our
host galaxy properties to those of other samples of galaxies. 

\begin{figure}[!hp]
\plottwo{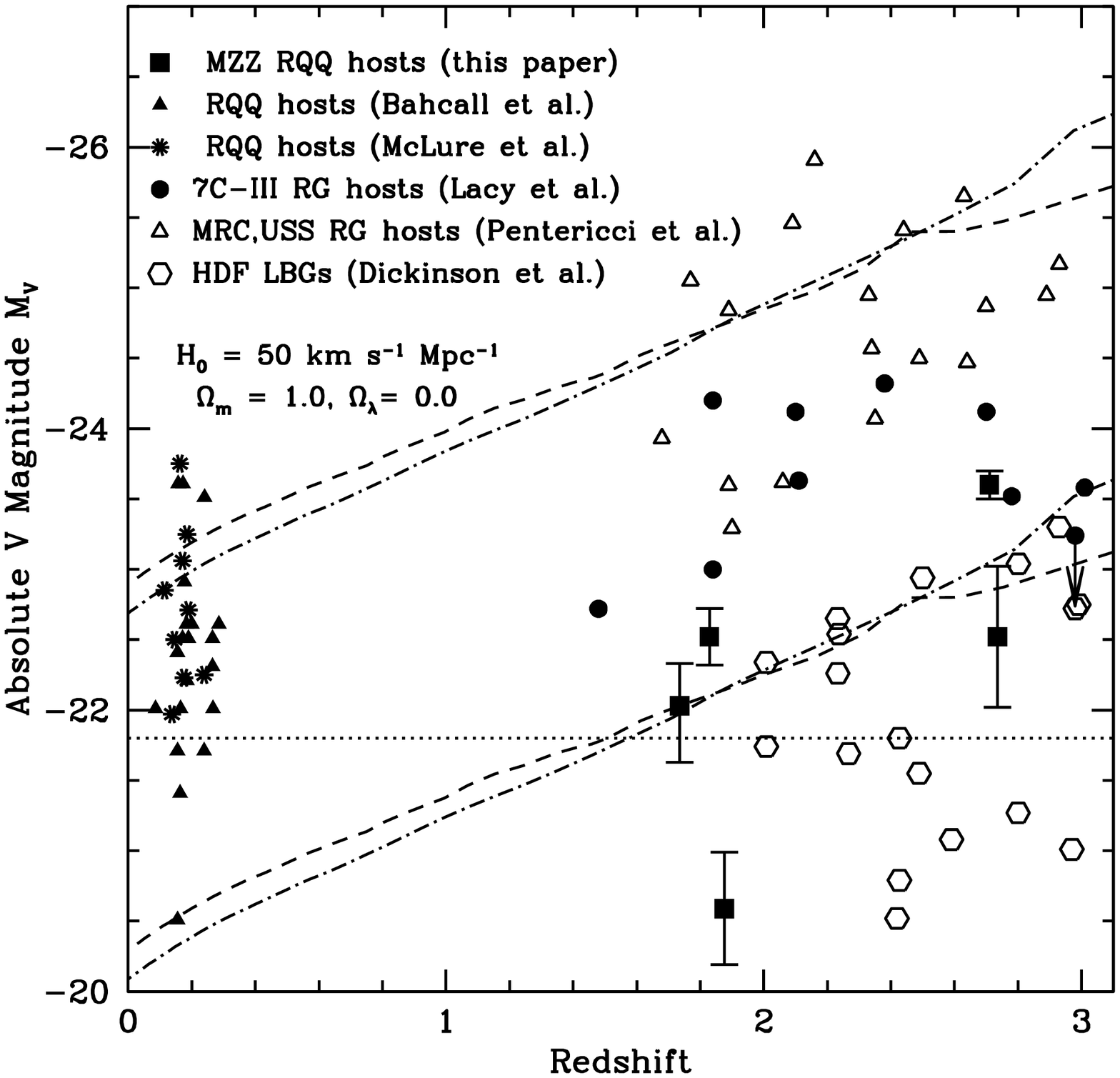}{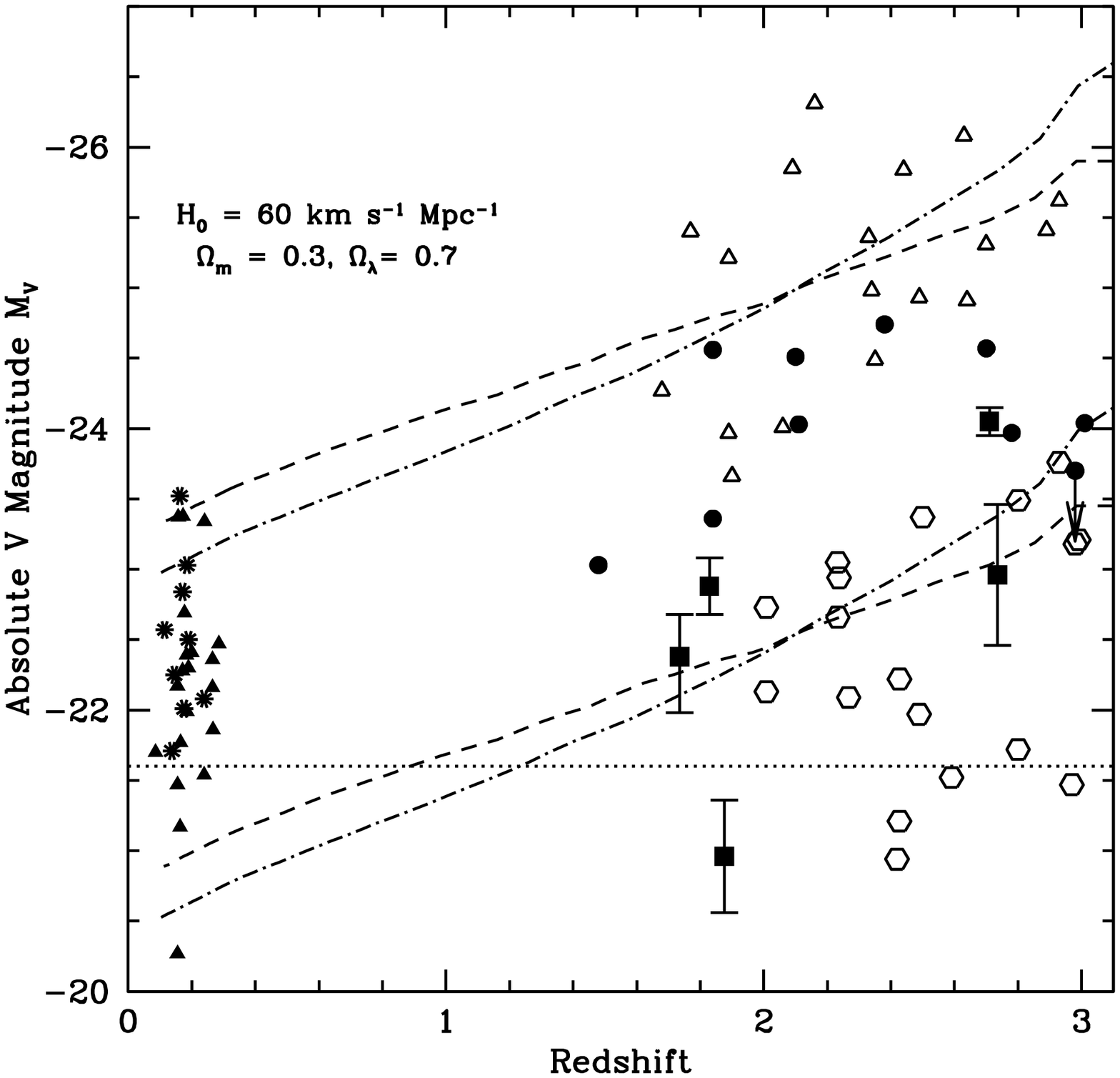}
\caption{
\label{mabsvfig}
The absolute rest-frame $V$ magnitudes versus redshift in two cosmologies
for our quasar hosts (squares) and
other samples of host galaxies: the low-$z$ quasar samples of Bahcall et al. (filled triangles)
and McLure et al. (stars), the 7C radio galaxy sample of Lacy et al. (filled circles), and the powerful
MRC and USS radio galaxy sample from Pentericci et al. (unfilled triangles). 
All magnitudes are total magnitudes. The horizontal dotted line indicates
present-day L$_*$. 
The dashed line represents the passive evolution of a model galaxy 
formed in an instantaneous burst of star formation at $z = 5$, while the
dot-dashed line represents a model with a 1 Gyr burst of star formation
ending at $z = 3$.  
These Bruzual \& Charlot (2000) models have been generated with a 
Salpeter IMF and an upper mass cutoff of 100 M$_\odot$.
The lower tracks for the instantaneous burst
are normalized to a total mass of 1.1 $\times$ 10$^{11}$ M$_\odot$
($\Omega_m$=1) or 2.0 $\times$ 10$^{11}$ M$_\odot$ ($\Omega_m$ = 0.3). The upper tracks 
are normalized to a total mass of 1.2 $\times$ 10$^{12}$ M$_\odot$ ($\Omega_m$=1) 
and 1.9 $\times$ 10$^{12}$ M$_\odot$ ($\Omega_m$ = 0.3).
}
\end{figure}

We plot in Fig. \ref{mabsvfig}, for two different 
cosmologies, the absolute rest-frame $V$ total magnitudes
of the 5 MZZ quasar hosts,
extrapolating our $z \sim 2.7$ quasar host $M_B$ values
to $M_V$ assuming the hosts have a spectral index of $-$2 (i.e. flat in F$_\lambda$), corresponding to a ($B$-$V$) of 0.7.  

We also correct to rest-frame $M_V$ and plot on these figures 
the Lyman break galaxy magnitudes from the Dickinson (2000)
sample, the powerful MRC/USS radio galaxies
of the Pentericci et al. (2000) sample,
and at low-$z$, the quasar host magnitudes from the Bahcall et al. (1997) 
and McLure et al. (1999)
samples. In addition, Lacy et al. (2000) have made ground-based $K'$
imaging studies of a $z \sim 2$ -- 3 sample of faint 7C-III radio galaxies 
for which we have both small and large aperture magnitudes, and we have
included these as well. 

We have used in all cases simple estimates of total magnitudes, applying 
corrections to the available aperture magnitudes of generally less than 
0.3 mags. 
For the HDF LBG galaxies, we measured the magnitudes directly from 
the image provided by Dickinson et al. (2000). As these galaxies have similar
compactness to our host galaxies, we used the same metric apertures used 
for our quasar hosts (i.e. 2\arcsec\ at $z = 1.8$ and $z = 2.7$, corresponding 
to $\sim$ 16 kpc), and made similar aperture corrections to total magnitude
as those used for the MZZ quasar hosts.
For the Pentericci et al. (2000) bright radio galaxy sample,
we have corrected their quoted 4\arcsec\ aperture magnitudes 
to total magnitudes assuming the hosts are ellipticals with scale sizes of 
10 kpc (corrections of $\sim$0.25 mags). For the 7C-III radio galaxy
sample of Lacy et al. (2000), we correct the quoted 64 kpc aperture magnitudes
to total, with the same assumptions (resultant corrections $\sim$0.15 mags). 
The low-$z$ quasar host magnitudes of Bahcall et al (1997) and McLure et al. (1999)
are discussed in more detail in section \ref{sec:lowzsim},
but do not require correction to total magnitude. 
All magnitudes were observed at close to rest-frame $B$ or $V$ and 
therefore the $K$-corrections are minor. 

The MZZ quasar host galaxy magnitudes at both $z$ ranges are
much more consistent with those of the Lyman break galaxies at
high-$z$ than with either sample of radio galaxies. 
At $z \sim 1.8$, our mean $M_V$ is about $-$21.5, while the 
radio galaxies have mean $M_V$ $\sim$ $-24.0$. At higher $z$, the 
difference is less, particularly between the less luminous 7C-III galaxies
and our hosts, but the means still differ by almost 2 magnitudes.
When compared to the Lyman break galaxies, the MZZ hosts span the
same range of magnitudes over this redshift range. 

Our derived magnitudes are similar to those of the low-$z$ quasars: slightly
less luminous in the high $\Omega_m$ cosmology, and comparable in
the low $\Omega_m$ cosmology.
However, the stellar masses corresponding to these absolute $V$ magnitudes
will be much less for the young, high-redshift galaxies than for the old
stellar populations probably associated with the low-$z$ galaxies. 
We have placed models of the passive evolution
of some simple stellar populations on the figures.

These models have been generated with the most recent of the Bruzual \& Charlot
population synthesis models (Bruzual \& Charlot 2000), using a 
Salpeter IMF and an upper mass cutoff of 100 M$_\odot$, with metallicity 
$Z = 0.02$ (Charlot 2000, priv. comm; further discussion of these models
can be found in Bruzual \& Charlot 1993; Liu, Charlot, and Graham, 
astro-ph/0004367). 
The dashed line represents a model
with an instantaneous burst of star formation at $z = 5$, while the
dot-dashed line represents a model with a 1 Gyr burst of star formation
ending at $z = 3$. The evolutionary tracks are roughly normalized
to the absolute $V$ magnitudes at $z = 1.8$ of (1) the MZZ quasar hosts
and 2) the powerful 
radio galaxies. This corresponds to total masses in {\it e.g.},
the instantaneous burst model, of 
(1)  1.1 $\times$ 10$^{11}$ M$_\odot$
($\Omega_m$=1) or 2.0 $\times$ 10$^{11}$ M$_\odot$ ($\Omega_m$ = 0.3) and (2) 
1.2 $\times$ 10$^{12}$ M$_\odot$ ($\Omega_m$=1)
or 1.9 $\times$ 10$^{12}$ M$_\odot$ ($\Omega_m$ = 0.3). 
The total stellar masses for the 1 Gyr burst models are about 20 -- 30\% smaller.
The degree of fading is about 2.2 -- 2.5 magnitudes in $V$ from $z=2.7$
to $z \sim 0$ and 1.5 -- 1.7 magnitudes from $z = 1.8$ to $z \sim 0$,
for the range of models and cosmologies considered, corresponding 
to (M/L) increases of $\sim$ 10 from $z = 2.7$ to $z \sim 0$. 
In other words, at $z = 0.2$, our mean quasar host would be
about 2 magnitudes fainter than the 
mean $z = 0.2$ quasar host from the McLure et al. and
Bahcall et al. samples if it evolved passively
from a short burst of star formation at $z = 3$ -- 5. 
This implies that the mean MZZ 
quasar host may have about one sixth the stellar mass of the
mean low-$z$ quasar host. 

Thus we find that the hosts of $z \sim 2$ -- 3 radio quiet quasars with
faint nuclei ($M_B$ $\sim$ $-$24 to $-$22) have magnitudes that
are consistent with their being drawn from the population
of Lyman break galaxies observed at similar redshifts. 

Our results are consistent with other recent NICMOS results on some brighter,
lensed quasars;
Rix et al. (1999) find that their sample of lensed $z \sim 2$ radio-quiet 
quasars (de-magnified $M_B$ $\sim$ $-$24 to $-$28)
had hosts with comparably faint magnitudes. 
In addition, first results from the large NICMOS quasar host survey of
Kukula et al. (2000) have been reported, giving similar or 
slightly brighter host magnitudes for
a sample of 5 $z \sim 2$ RQQs with nuclear $M_B \sim$ $-$24.

These magnitudes
are not consistent, however, with those of the sample of
radio galaxies at similar redshifts nor with these high-$z$ 
quasars passively evolving into the hosts of the low-$z$ quasars. 

\subsection{Host galaxy morphologies compared}

We also compare the morphologies of our quasar hosts with those
of the other samples of high-$z$ objects for which there are good
estimates of scale sizes in the rest-frame $\sim V$ or $B$. 
Though our determinations of scale sizes are very approximate, 
in no cases did we find evidence for a half-light radius more than 4 kpc
($\sim$0\farcs5 at these redshifts). 
These scale sizes
are more consistent with those of the Lyman break galaxies than with those
of the high-$z$ radio galaxies or low-$z$ quasar hosts, and our simulations
have strengthened this result. 

The $z = 2 $ -- 3 HDF-N
Lyman break galaxies are found to be quite compact in the rest-frame
UV WFPC2 imaging, with scale sizes of $\sim$0\farcs2--0\farcs3
(Giavalisco et al. 1996;
Lowenthal et al. 1997). The NICMOS F160W imaging, most comparable to 
our studies of the MZZ quasars, revealed similar compactness in the
LBG sample, with identical or even more compact half-light radii
than those measured in the optical (Dickinson 2000). 
On the other hand, the 
low-$z$ quasar hosts 
are found to have half-light radii of 8 -- 15 kpc in the
rest-frame $V$ WFPC2 imaging, 
consistent with their being relaxed giant ellipticals or evolved
disk galaxies (McLure et al. 1999; Bahcall et al. 1997).

Comparison to the morphologies of the high-$z$ radio galaxies is less simple,
as there may be some dependence of the half-light radii
on the luminosity of the radio sources (e.g. Roche, Eales \& Rawlings 1998; 
Lacy et al. 2000). Generally, however, at $z \lesssim 1.5$, 
the galaxies (with a range in radio 
luminosities) are found to be well-fit by $r^{1 \over 4}$ laws, with
half-light radii in the range of 5 -- 11 kpc (Roche et al. 1998; Best et al.
1998; McLure \& Dunlop 2000; Lacy et al. 2000).
Pentericci et al. (2000), however, indicate that the powerful radio
galaxies in their sample at
$z > 2$ have scale sizes 3 times less than those at $z \sim 1$, while
Lacy et al. (2000) find less evolution in scale size.
While the best determinations
of scale size at rest-frame $V$ are these NICMOS observations of
the MRC/USS sample of Pentericci et al. (2000), their sample 
is fairly small and reveals a large range in scale size. In 5 cases, they were
able to fit de Vaucouleurs profiles, and determined $r_{1 \over 2}$
values ranging from 0\farcs2 to 1\farcs6. 
In Lacy et al. (2000), a summary of the available morphological data in
the literature (including the measurements of Pentericci et al.),
plus the results of ground-based scale size determinations
on the 7C-III sample, indicate that at $z \sim 2$ -- 3 the mean scale
size of high $z$ radio galaxies
is $\sim$0\farcs6, with the same large range found in the MRC/USS sample.
Thus, although some of the high-$z$ radio galaxies have scale sizes as
compact as the LBGs or the MZZ quasar hosts, the mean $z \sim 2 $ --3
radio galaxy is probably still more extended.

With such a small number of objects we cannot effectively
judge how significantly ``disturbed'' the morphologies in this sample 
are in comparison to other samples of high-$z$ galaxies. 
Three of our 5 quasars
have discrete companion objects or a compact component within 10 kpc of the quasar and 
otherwise seem to have regular morphologies.
Overall, this level of ``disturbance'' in the sample
is consistent with the complex morphologies seen in the Lyman break and
radio galaxy samples. 
However, even the low-$z$ quasar hosts have some 
discrete residual components and tidal tails 
that are visible after subtraction of the main galaxy component (e.g. 
McLure et al.  1999), and an unusually high incidence of nearby
companions is also noted (Bahcall et al. 1997).

\subsection{Black hole masses, quasar luminosities, and quasar host properties}

We also wish to investigate the relationship between
the quasar host properties, the luminosities of the quasar nuclei,
and the masses of the nuclear black holes. 
In particular, a number of results have lent support to the idea that the
AGN luminosity and the mass of the host galaxy may be linked.
Studies of nearby
spheroids indicate that nearly every present-day stellar spheroid contains
a massive black hole (BH), and that the masses of the black holes
and the luminosities of the bulges (and therefore by inference
the masses of the bulges) are linearly correlated
(Kormendy \& Richstone 1996;
Magorrian et al. 1998; van der Marel 1999). 
The stellar velocity dispersions of these bulges may provide a more
direct measure of their masses, as evidenced by the even tighter
correlation of velocity dispersion with black hole mass
(Ferrarese \& Merritt 2000; Gebhardt et al. 2000a). 
The BH masses determined 
from this velocity dispersion correlation are consistent with those
determined from reverberation mapping of AGN (Gebhardt et al. 2000b).
All of these results imply a close link between the formation of the 
bulges of the galaxies and the formation of the black holes found within them. 

If more massive black holes
translate into more powerful quasar nuclei,
we might then expect that the brighter AGNs (including both
brighter radio sources and quasars of greater nuclear luminosity)
would have the most luminous and most massive hosts.
In fact, there is an upper bound to the luminosity of the low-$z$ quasars
that is consistent with Eddington-limited
accretion onto supermassive black holes with masses given by
the local $M_{BH}$ {\it vs.} $M_{bulge}$ relation
(McLeod, Rieke, \& Storrie-Lombardi 1999; McLure et al. 1999; Laor 1998).

To make an estimate of the masses of the black holes
associated with these MZZ quasars, we  
can apply the local $M_{BH}$ {\it vs.} $M_{bulge}$ relation to 
our host galaxies. We have used the van der Marel (1999) normalization
of this relationship, and have made corrections for the much lower
M/L ratio at the redshifts of our objects than at $z = 0$ (estimated from the 
passive evolution models discussed earlier). This results in 
black hole masses mostly around 1 $\times$ 10$^{8}$ M$_\odot$,
but ranging from 0.3 -- 2 $\times$ 10$^{8}$ M$_\odot$. These values
are given in Table \ref{bhtab}. 
We then use these derived black hole masses and the $M_B$ nuclear magnitudes
of the quasars to estimate the ratio of the quasar luminosity 
to the Eddington luminosity, making
a bolometric correction to 
the $B$ band of a factor of 12 (Elvis et al. 1994).  
We find accretion rates that vary from 0.2 to 1.4 times the Eddington
limit, with most at 70\% of the Eddington accretion rate (also 
in Table \ref{bhtab}). 
This is consistent with schemes in which high-$z$ radio-quiet 
AGN are associated with moderately massive black holes
and accrete at close to the Eddington rate, while luminous radio-loud
objects are believed to be accreting at sub-Eddington rates and to be 
associated only with the most massive black holes ($>$10$^{9}$ M$_{\odot}$)
(Rawlings \& Saunders 1991; McLure et al. 1999; Willot et al. 1999; 
Lacy, Ridgway \& Trentham 2000).
In contrast, Rix et al. (1999) found comparable host magnitudes
for a much brighter set of quasars, and hence found accretion rates
that were super-Eddington. 

Another way of estimating the black hole masses in AGN is by using the 
velocity dispersion of the $H\beta$ emitting clouds in the broad-line region
(BLR), (as measured by the $H\beta$ FWHM), in combination with 
some estimate of the size of the BLR emitting region, 
and assuming that the dispersion is from virialized cloud motion
(e.g. Laor 1998; Wandel et al. 1999).
Wandel et al. compare reverberation mapping estimates of the black hole
masses in a sample of Seyferts and quasars to estimates from the $H\beta$
FHWM, and found they correlated well. 
Laor (1998) found that the $H\beta$-derived BH masses in the Bahcall 
et al. quasar sample correlated well with the black hole masses derived from
the Magorrian et al. (1997) $M_{bulge}$ / $M_{BH}$ relation.
From the identification spectra for the MZZ quasars (Vitelli et al. 1992),
we can estimate very rough CIV FWHMs for 4 of the quasars 
(in the range 6500 to 7500 km s$^{-1}$),
and use these as representative of the $H\beta$ widths. Laor (1998)
estimates the radius of the BLR as proportional to the square root of
the bolometric luminosity, and adopting his formulation and our previous
estimates of the bolometric nuclear luminosity we find 
black hole masses for these quasars of about 3 to 8 $\times$ 10$^{8}$
M$_\odot$. The derived masses are roughly proportional to those we derived
from the van der Marel (1999) relation but are about a factor of 4 larger.
However, both Wandel et al. and Laor found that their correlations
required separate calibration, and as we have used a different 
line, and the available spectra are insufficient to make reliable estimates
of the line FWHMs, this discrepancy is not surprising. 

\subsection{Implications}

Observations of galaxies at high-redshift to date have focussed largely
on two very distinct populations: powerful radio-loud AGN (chiefly
radio galaxies, but including radio-loud quasars), and
normal star-forming galaxies (the Lyman break population). One 
of the major goals of our investigation is to better understand
the relationship of the hosts of typical radio-quiet AGN
to these other populations.

We have found that the radio-quiet quasar hosts are similar
to the Lyman break galaxies in terms of rest-frame optical
luminosities and sizes, but are considerably less luminous
and smaller than radio galaxies at similar redshifts. These results
have some interesting implications. We will consider the comparisons
to the radio-loud AGN hosts first, and then briefly discuss
the Lyman break galaxies.

\subsubsection{Evolution of quasar and radio galaxy hosts}

Our results reveal an interesting discrepancy between the evolution of
the radio-quiet and radio-loud host galaxies. At low-$z$, luminous
radio-quiet quasars are found primarily in massive early-type galaxies
with luminosities of
several times a present-day $L_*$ galaxy (e.g. Bahcall et al. 1997;
McLure et al. 1999; McLeod et al. 1999), 
and have properties that are comparable to those of the radio galaxies.
McLure et al. have compared carefully selected samples
of radio-quiet quasars, radio-loud quasars, and radio galaxies
at $z \sim 0.2$ and shown their host properties are all similar (with
the radio-loud objects in slightly brighter hosts that those of the
radio-quiet). 
However, despite being nearly indistinguishable from radio galaxy and radio
loud-quasar hosts at low $z$, at high $z$ the radio-quiet quasar hosts 
are several magnitudes fainter than the radio galaxies. 

The cosmic evolution of the population of powerful radio galaxies has
been well-documented over the past decade (e.g.  R\"ottgering, Best,
\& Lehnert 1999). 
Radio galaxies (over a wide range in radio power) are found in relaxed,
massive elliptical hosts up to fairly high redshift ($z \sim 2.5$) and
have $K$ magnitudes with low dispersion, 
consistent with their having formed at much higher $z$ and
then subsequently passively evolved into 
present-day giant ellipticals (Lilly \& Longair 1984, Lilly 1989,
Rigler et al. 1992, Eales et al. 1996, Best et al. 1998, McCarthy 1999). 
Recent work has indicated that at $z \sim 2.5$ or higher, radio galaxies
have more unsettled morphologies (Pentericci et al. 1999, van Breugel et al.
1999), and there is a larger spread in absolute magnitude (Lacy et al. 2000),
indicating that at $z \sim 3$ -- 5 we may be reaching the epoch of radio galaxy
formation. 
Much less is known about the evolution
of the population of the hosts of radio-loud quasars, but the available data
paint a broadly similar picture (e.g. Lehnert et al. 1992, 1999; 
Ridgway \& Stockton 1997; McLure et al. 1999).

While the most powerful radio galaxies are extremely rare objects, 
the lower-luminosity radio sources have much higher space densities. 
For H$_0$ = 50 km s$^{-1}$
Mpc$^{-1}$ and $\Omega$ = 1,
3CR radio galaxies at $z$ = 2.5
have a co-moving space density of only 0.2 Gpc$^{-3} \Delta$log$P_{rad}^{-1}$
while the fainter 6C (Eales et al. 1997), 7C (Lacy et al. 1999; Willott 2000),
and MRC (McCarthy et al. 1996) radio
galaxies have space densities roughly $10^2$ -- $10^3$ times larger (Dunlop
\& Peacock 1990). These values
can be compared to the present-day space-density of first-ranked cluster
galaxies (roughly 5000 Gpc$^{-3}$, Bahcall \& Cen 1993). Allowing for
a short lifetime for the radio galaxy phase,
the evolved descendants of radio sources over a range in luminosities
would account for a modest fraction of the first-ranked cluster 
elliptical galaxies.

In contrast, however, radio-quiet quasars of the
luminosity we have studied in this paper are far more common, and therefore
are more plausible progenitors of typical present-day early-type galaxies.
Radio-quiet quasars with $M_B \leq$ $-$23 have co-moving space densities
at z$\sim$2 of $\sim$2$\times$10$^4$ Gpc$^{-3}$
(Hartwick \& Schade 1990). Now, for a quasar lifetime on the order of the
Eddington growth-time (a few \% of the
Hubble time at z = 2), the implied space density of the present-day
descendants is about 10$^6$ Gpc$^{-3}$, which is comparable to the space
density of $L_*$ E's and S0's (e.g. Fukugita, Hogan, \&
Peebles 1998). This identification is quite consistent with
the correlation between the masses of supermassive black holes and
the spheroids in which they live today. A quasar with $M_B$ = $-$24
powered by accretion at the Eddington rate requires $M_{BH} \sim$
2 $\times$ 10$^8$ $M_{\odot}$, and this supermassive black hole
would live today in a spheroid with a mass of 
about 5 $\times$ 10$^{10}$ $M_{\odot}$ and a 
V-band luminosity of about 1.5 $\times$
10$^{10}$ $L_{\odot}$ $\sim$ 0.4 $L_*$.

Though inconsistent with passive evolution like that observed
in the radio galaxies, our finding L$_*$ hosts
at $z \sim 2$ -- 3 (and the similar results reported by Rix et al. [1999]) 
are reassuring for standard hierarchical models.  
The types of quasar
hosts we have imaged can easily be the precursors of typical present-day
bulge-dominated $L_*$ galaxies, provided that they continue to grow
somewhat through mergers from $z \sim$ 2 to 3 until the present day.
Our typical high-$z$ host might need to accrete a factor of few in mass
in order to reach the magnitude of a $z = 0$ $L_*$ galaxy.

\begin{figure}[!p]
\plottwo{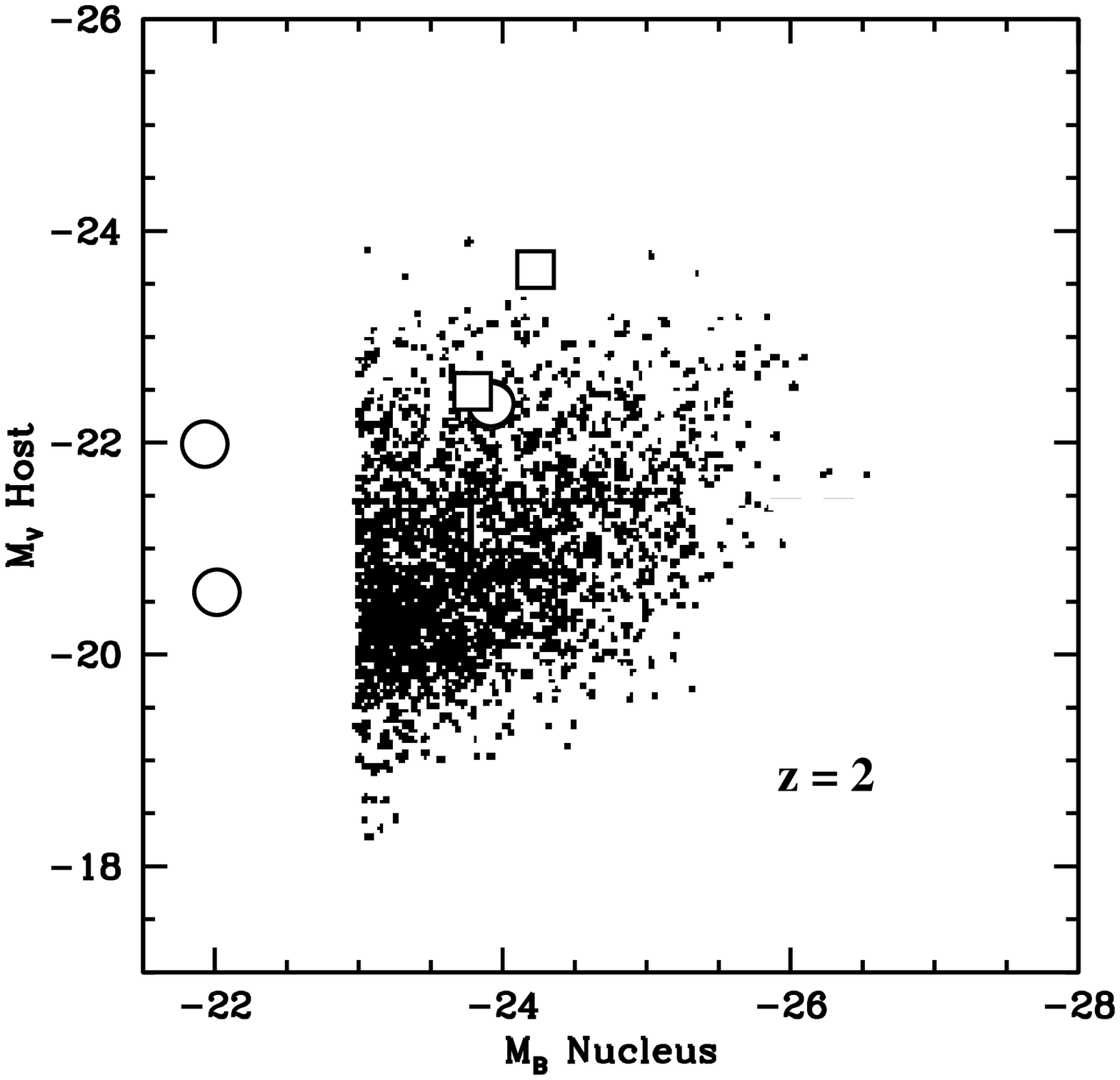}{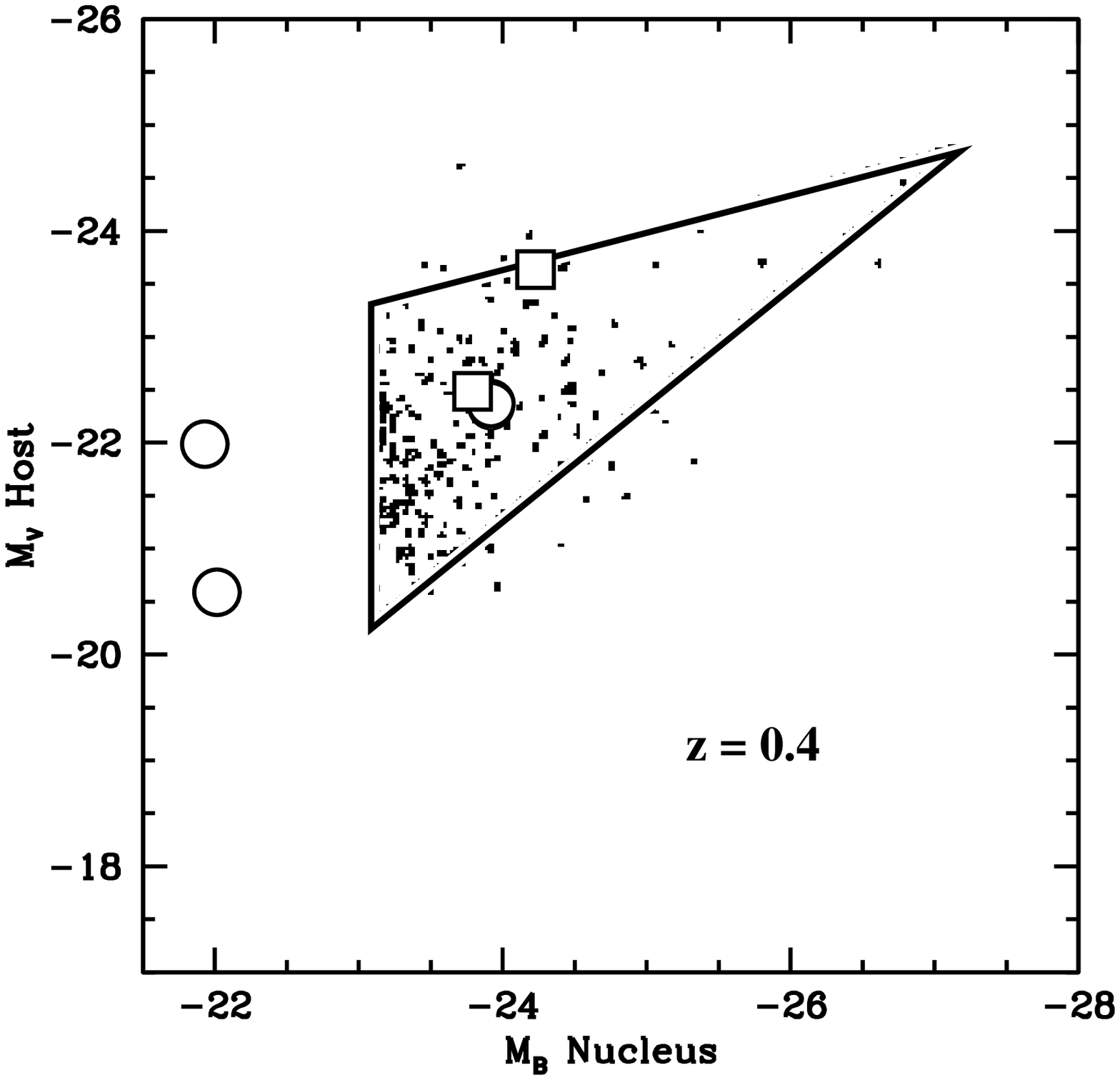}

\caption{Results for the MZZ host galaxy magnitudes plotted as large
unfilled symbols over the theoretical predictions (black dots)
from Kauffmann \& Haehnelt (2000) (adopted from their figure 12).
The squares are our $z \sim 2.7$ quasar hosts, the circles
are the $z \sim 1.8$ quasars, placed on both the low-$z$ and
$z \sim 2$ planes. 
The triangular region encompasses the low-$z$ quasar hosts of McLeod et al. (1999). 
\label{kaufffig}
}
\end{figure}

Indeed, these results agree fairly well with the specific predictions
of the hierarchical galaxy formation models of Kauffmann \& Haehnelt (2000),
in which they have addressed both the formation of bulges 
and the formation and fuelling of their associated black holes. 
They predict median host luminosities that are somewhat below
present-day L$_*$ for quasars at $z=2$ (and even fainter at $z=3$),
for quasars with the nuclear magnitudes of our sample. In Figure \ref{kaufffig}
we show our data superimposed on their models. Our galaxies span the range
of their model results, though our average host is brighter than the average of the
model galaxies shown. 
In their model, these hosts are still
undergoing major mergers which would allow them to evolve into the present
day $L_*$ or several $L_*$ galaxies. 
These models are also able to reproduce the
tight correlation of bulge velocity dispersion to black hole mass 
(Haehnelt \& Kauffmann 2000).

As the typical quasar lifetime is probably short, in observing 
active quasars we may be selecting quite
different samples of objects at $z \sim 2$ -- 3, the peak of the
quasar number density, from those at low-$z$. 
At low-$z$, where the number density of quasars
has decreased dramatically, 
the hosts of most quasars are rarer and somewhat more massive
than the more common $L_*$ galaxies that are the possible descendants of our 
high-$z$ quasar hosts. 
These high $z$ hosts may evolve therefore into  
more common, less active low-$z$ counterparts, such as Seyferts and 
quiescent $L_*$ spheroids.  
This would make their evolution similar to those of
the field ellipticals and Lyman break galaxies, 
(Dickinson et al. 1999; Steidel et al. 1999), whereas the most
massive galaxies like the radio galaxies will undergo a very different
evolution, associated with an early formation epoch. 

\subsubsection{Relation to the Lyman break galaxy population}

The co-moving space density of the Lyman break galaxies
(near the turnover of their UV luminosity function) is roughly
10$^6$ Gpc$^{-3}$ (e.g. Dickinson 1998). If the hosts of radio-quiet
quasars are drawn from the Lyman break population, then we conclude
that only a few percent of these galaxies need host a radio-quiet quasar
at any given time. If there is a population of unbeamed (``type 2'') quasars
at high-$z$ (analogous to the Seyfert 2 galaxies at low-$z$) then
the total fraction of the Lyman break galaxies hosting an AGN might
be several times larger. These relative numbers are consistent with
a simple model in which the lifetime of the Lyman break phase
(high star-formation rates) is roughly 10$^9$ years (cf. Ferguson,
Dickinson, \& Williams 2000) and the AGN phase is roughly 10$^8$ years
(comparable to the characteristic Eddington growth time).

Could a significant fraction (a few percent) of the known
Lyman break population population be the ``quasar 2's''?
By hypothesis, these would be objects in which the quasar
is hidden from direct view along our line-of-sight. By analogy
to Seyfert 2 galaxies in the local universe (e.g. Heckman
et al. 1997; Gonzalez Delgado, Heckman \& Leitherer 2000) 
the rest-frame UV light
in these galaxies would be dominated by light from young stars,
and so these could naturally enter the color-selected
Lyman break samples. If they are like local type 2 Seyferts, they
could be distinguished from the
majority of ``normal'' Lyman break galaxies because they
would show relatively strong but rather narrow 
($\leq$ 10$^3$ km s$^{-1}$) nebular emission-lines. 
The Ly$\alpha$ line would normally be the strongest, but lines
from highly ionized species like HeII $\lambda$1640, CIV $\lambda$1550,
and NV $\lambda$1240
should also be present
with strengths of roughly 5 to 20\% of Ly$\alpha$. This picture
appears to be quite consistent with the spectroscopic properties
of the Lyman break galaxies. C. Steidel (private communication)
finds that $\sim$ 1.5\% of spectroscopically confirmed Lyman break
galaxies at z$\sim$3 are obvious narrow-lined AGNs with
Seyfert-2-like spectra. This is a lower bound to the actual fraction,
since the weak NV, CIV, and HeII lines could be missed in many
other cases.

An obvious way to test the similarity between the Lyman break galaxies
and the hosts of high-redshift AGN would be to obtain rest-frame UV images
of the quasar hosts in the present sample. Our on-going HST WFPC2
imaging program will make this test possible.

\section{Summary}

We have reported the results of the analysis of $HST$ NICMOS images of
5 faint (M$_B \sim$ $-$23) radio-quiet quasars at redshifts ($z\sim$ 2 to 3)
near the peak of the quasar epoch in the early universe. Our work
complements the analysis reported by Rix et al. (1999)
of six luminous ($M_B \sim$ $-$26) gravitationally-lensed radio-quiet
quasars at similar redshifts. 

While the samples are still modest in size, several conclusions can already
be drawn:

\begin{itemize}

\item
Typical radio-quiet quasars at $z \sim$ 2 to 3 are hosted by galaxies with
rest-frame absolute visual magnitudes similar to present-day
$L_*$ galaxies ($M_V$ = $-$20 to $-$23 $\equiv$ $M_{*,V}\pm$1.5 mag).

\item
As such, they are much fainter than radio galaxies at the same redshift
(typically by $\sim$ 2 magnitudes). 

\item
These host galaxies are comparable to or less luminous than the
hosts of similarly-powerful low-$z$ radio-quiet quasars.
Since the luminosity-weighted mean age of the stellar population
in the high-$z$ hosts is almost certainly younger than that of
the low-$z$ hosts, the difference in stellar {\it mass} will be even more
pronounced, by perhaps a factor of 6.
The high-$z$ hosts are also more compact than the
low-$z$ hosts (half-light radii typically $<$4 kpc {\it vs.} 8 kpc).

\item
The rest-frame-visual luminosities and sizes of the radio-quiet
quasar hosts are roughly similar to those of the Lyman-break galaxies
at similar redshifts.
Thus, the Lyman-break population could represent the parent
population of typical radio-quiet quasars. Our Cycle 8 HST WFPC2 observing
program will determine whether this similarity extends into
the rest-frame UV.

\end{itemize}

The potential implications of these results are quite significant.
First, they imply that the well-studied cosmic evolution of the
hosts of the very radio-loud AGN population (radio galaxies and radio-loud
quasars)
is evidently not representative of the much more numerous radio-quiet
population. Second, by assuming either 
that the ratio of $L_{Q}/M_{BH}$ is roughly independent of
redshift or that $M_{BH} \propto M_{bulge,z=0}$, it follows that
these quasars are associated with black holes with masses
of about 10$^{8}$ M$_{\odot}$. Therefore, 
supermassive black holes must form before their host galaxies are fully
assembled. This agrees qualitatively with the idea of the hierarchical assembly
of massive galaxies at late epochs. Indeed, as already pointed out
by Rix et al. (1999) and Ridgway et al. (1999), the
observations agree well with the recent theoretical predictions of 
Kauffmann \& Haehnelt (2000). We might expect that the average low-$z$
counterparts of these high-$z$ quasar hosts may be quiescent bulge-dominated
systems rather than the massive galaxies associated with the low-$z$
active quasar population. 

\acknowledgments {
We thank Eddie Bergeron for advice in the NICMOS data calibration.
We thank Mark Dickinson for use of the HDF-N Lyman break
galaxy sample and images, Pat McCarthy for use of the NICMOS images of 
the MRC/USS radio galaxy sample, and Mark Lacy for helpful discussions.
Support for this work was provided by NASA through grant number
GO-07864.01
from the Space Telescope Science Institute, which is operated by AURA,
Inc., under NASA contract NAS5-26555.
}

\clearpage
	\begin{deluxetable}{l c c c c c c c c}
	\tablecaption{Properties of sample and observational log \label{obslog}}
\tabletypesize{\footnotesize}
	\tablehead{
	\colhead{Name} & \colhead{RA} & \colhead{DEC} & \colhead{Redshift} & 
	\colhead{B} & \colhead{Filter} & \colhead{$\lambda_0$\tablenotemark{a}} & \colhead{Exposure Time} & \colhead{$\sigma_{\rm {sky}}$\tablenotemark{b}} \\
	\colhead{} & \colhead{J2000} &\colhead{J2000} &\colhead{} & \colhead{(Nuclear)} &\colhead{} &\colhead{(\AA)} &\colhead{(seconds)} &\colhead{($\mu$Jy arcsec$^{-2}$)}
	} 
	\startdata
	MZZ 9744 & 03 13 38.33 & $-$55 21 37.0 & 2.735 & 21.9 & F160W & 4284 & 11007 & 0.48 \\ 
	MZZ 9592 & 03 14 04.94 & $-$55 20 50.9 & 2.710 & 21.9 & F160W & 4312 & 12543 & 0.48 \\
	MZZ 1558 & 03 14 51.48 & $-$54 57 14.5 & 1.829 & 21.6 & F165M & 5832 & \phn9919 & 0.97 \\ 
	MZZ 11408 & 03 15 34.10 & $-$55 30 04.8 & 1.735 & 22.0 & F165M & 6033 & \phn9983 & 1.10 \\
	MZZ 4935  & 03 16 36.25 & $-$55 09 32.2 & 1.876 & 21.9 & F165M & 5737 & 18943 & 0.75 \\ 
	\enddata
	\tablenotetext{a} {Mean rest wavelength of observation.}
	\tablenotetext{b} {1$\sigma$ scatter per pixel in surface brightness 
from sky noise (1 pixel = 0\farcs0375). }
	\end{deluxetable}

	\begin{deluxetable}{l c c c l l c c}
	\tablecaption{Host galaxy photometry \label{maghosttab}}
\tabletypesize{\footnotesize}
	\tablehead{
	\colhead{Name} &\colhead{Redshift} &
	\colhead{$M_B$} &\colhead{Aperture}& \colhead{$H$ Host flux}&
	\colhead{$M_{\rm host}$} & \colhead{Host} & \colhead{Total}\\
	\colhead{}&\colhead{} &\colhead{(Nucleus)}
	 &\colhead{radius} &\colhead{($\mu$Jy)} &\colhead{} &
	 \colhead{luminosity} & \colhead{$M_{\rm host}$\tablenotemark{a}}
	}
	\startdata

	MZZ 9744 &  2.735 & $-$23.8 & 0\farcs64 & 1.08$\pm$0.4 & $-$21.3 (B) $\pm$ 0.5 & 1.3 $L_*^B$ & \\ 
	\nodata &  \nodata & \nodata & 1\farcs01 & 1.32$\pm$0.4 & $-$21.5 (B) $\pm$ 0.5 & 1.6 $L_*^B$ & $-$21.8 (B)\\ 
	MZZ 9592 &  2.710 &$-$24.2 & 0\farcs64 & 3.21$\pm$0.2 & $-$22.5 (B) $\pm$ 0.1 & 3.9 $L_*^B$ & \\
	\nodata &  \nodata  & \nodata & 1\farcs01 & 3.37$\pm$0.2 & $-$22.5 (B) $\pm$ 0.1 & 4.0 $L_*^B$ & $-$22.9 (B) \\
	MZZ 1558 &  1.829 & $-$23.8 & 0\farcs64 & 2.44$^{+0.4}_{-0.5}$ & $-$21.6 (V) $\pm$ 0.2 & 0.8 $L_*^V$ & \\
	\nodata &  \nodata & \nodata & 1\farcs01 & 3.27$\pm$0.5 & $-$21.9 (V) $\pm$ 0.2 & 1.1 $L_*^V$ & $-$22.5 (V) \\
	MZZ 11408 & 1.735 & $-$21.9 & 0\farcs64 & 2.63$^{+0.3}_{-0.7}$ & $-$21.5 (V) $^{-0.3}_{+0.4}$ & 0.8 $L_*^V$ & \\
	\nodata & \nodata & \nodata & 1\farcs01 & 3.12$^{+0.3}_{-0.7}$ & $-$21.7 (V) $^{-0.3}_{+0.4}$ & 0.9 $L_*^V$ & $-$22.0 (V)\\
	MZZ 4935  & 1.876 & $-$22.0 & 0\farcs64 & 0.58$\pm$0.2 & $-$20.1 (V) $\pm$0.4 & 0.2 $L_*^V$ & \\
	\nodata  & \nodata &  \nodata  & 1\farcs01 & 1.01$\pm$0.2 & $-$20.2 (V) $\pm$0.4 & 0.2 $L_*^V$ & $-$20.6 (V) \\
	\enddata
\tablenotetext{a}{This approximate ``total'' magnitude is estimated from the measured
$M_{\rm host}$ within 2\arcsec\ by adding the appropriate subtraction 
corrections from the HDF LBG models
(Table \ref{modelmagtab}) and the aperture corrections
given in Table \ref{proffittab}. }
	\end{deluxetable}

	\begin{deluxetable}{l c c c c }
	\tablecaption{Quasar host scale sizes \label{proffittab}}
	\tablehead{
	\colhead{Name} &\colhead{Redshift} &
	\colhead{$r_{1\over2}$\tablenotemark{a}} &
	\colhead{model type:$r_{1 \over 2}$\tablenotemark{b}}& 
	\colhead{Aperture Correction\tablenotemark{c}} \\
	\colhead{} & \colhead{} & \colhead{} & \colhead{} & 
		\colhead{($\Delta$ mags)} \\
	\colhead{(1)} & \colhead{(2)} & \colhead{(3)} & \colhead{(4)} 
	& \colhead{(5)} 
	}

\startdata

	MZZ 9744 &  2.735 & 0\farcs24 & \nodata & 0.1 \\
	MZZ 9592 &  2.710 & 0\farcs31 & disk: 0\farcs26  -- 0\farcs31 & 0.1\\
	\nodata &  \nodata & \nodata & bulge: 0\farcs11 -- 0\farcs18 & \nodata\\
	MZZ 1558 &  1.829 & 0\farcs39 & \nodata & 0.3 \\
	MZZ 11408 & 1.735 & 0\farcs25 & disk: 0\farcs24 -- 0\farcs22 & 0.2 \\
	\nodata & \nodata & \nodata & bulge: 0\farcs14 -- 0\farcs37 & \nodata \\
	MZZ 4935  & 1.876 & 0\farcs25 & \nodata &  0.2 \\

\enddata

\tablenotetext{a}{Half-light radius determined directly from enclosed flux 
profile of PSF-subtracted residual host. This determination assumes the inner
0\farcs2 is flat. }
\tablenotetext{b}{Range of $r_{1 \over 2}$ determined by fitting (where
 possible) disk and bulge profiles to the 1D radial
profile of the subtracted host.}
\tablenotetext{c}{Change in magnitude 
due to flux excluded from a 2\arcsec\ aperture, 
for a host galaxy with the given $r_{1 \over 2}$.}
	\end{deluxetable}

	\begin{deluxetable}{l c c c c c c c}
	\tablecaption{Quasar nuclear properties \label{tablenuc}}
\tabletypesize{\footnotesize}
	\tablehead{
	\colhead{Name} &\colhead{Redshift} & \colhead{Nuclear Flux($H$)\tablenotemark{a}} &
	\colhead{$M_B$(Nucleus)} & \colhead{$\alpha$\tablenotemark{b}} &
 \colhead{Nuclear}\\
	\colhead{}&\colhead{} &\colhead{($\mu$Jy)} &\colhead{} &\colhead{} 
& \colhead{Fraction}\tablenotemark{c}
	}
	\startdata

	MZZ 9744 &  2.735 & 10.6 $\pm$ 0.5 & $-$23.8 & $+$0.6 & 89\%\\
	MZZ 9592 &  2.710 & 15.9 $\pm$ 0.5 & $-$24.2 & $-$0.6 & 83\% \\
	MZZ 1558 &  1.829 & 24.5 $\pm$ 0.8 & $-$23.8 & $-$0.7 & 90\%\\
	MZZ 11408 & 1.735 & 3.9 $\pm$ 0.2 & $-$21.9 & $+$0.2 & 56\% \\
	MZZ 4935  & 1.876 & 3.4 $\pm$ 0.2 & $-$22.0 & $-$0.3 & 77\%\\
	\enddata
	\tablenotetext{a}{Nuclear flux derived from PSF subtraction fit,
includes all NICMOS PSF energy enclosed within a 5\farcs7 aperture}
	\tablenotetext{b}{The spectral index 
	$\alpha$ ($f_{\nu} \propto \nu^{\alpha}$),
	from observed $B$ to $H$. }
	\tablenotetext{c}{Percentage of total flux due to the nucleus.}

	\end{deluxetable}

\begin{deluxetable}{l c c c c c c}
\tablecaption{Summary of model results: magnitude corrections \label{modelmagtab}}
\tablehead{
\colhead{Model Type\tablenotemark{a}} & 
\multicolumn{2}{c}{Model Nucleus\tablenotemark{b}} &
\colhead{Quasars\tablenotemark{c}} & 
\multicolumn{3}{c}{$\Delta$ Magnitudes\tablenotemark{d}}\\
\cline{2-3} \cline{5-7} \\
\colhead{} & \colhead{$H$ Flux($\mu$Jy)} & \colhead{$M_B$\tablenotemark{e}} & 
 \colhead{} &
\colhead{HDF galaxies} & \colhead{Radio galaxies} &
 \colhead{Ellipticals/Disks} \\
\colhead{(1)} & \colhead{(2)} & \colhead{(3)} & \colhead{(4)} & \colhead{(5)} &
\colhead{(6)} & \colhead{(7)}
}

\startdata
$z = 1.8$ low &  \phn3.7 & $-$21.9 & MZZ 4935, MZZ11408 & 0.13 & 0.33 & 0.11\\
$z = 1.8$ medium &  14.5 & $-$23.4 & --- & 0.25 & 0.32 & 0.14\\
$z = 1.8$ high &  25.3 & $-$24.0 & MZZ 1558 & 0.32 & 0.32 & 0.25\\
\\
$z = 2.7$ low &  \phn5.0 & $-$22.9 & --- & 0.13 & 0.44 & 0.18 \\
$z = 2.7$ medium & 10.6 & $-$23.7 & MZZ 9744 & 0.16 & 0.45 & 0.35 \\
$z = 2.7$ high & 15.9 & $-$24.2 & MZZ 9592 & 0.25 & 0.46 & 0.43 \\
\enddata
\tablecomments{Here we summarize the results of the models discussed in the
text, in which we have created artificial quasars with a range of model
host galaxies and a range of nuclear brightnesses, and determined how much
flux was lost by the process of PSF-subtraction.}
\tablenotetext{a}{The redshift of the model, and whether a nucleus of
low, medium or high brightness was used.}
\tablenotetext{b}{The $H$ flux and absolute B magnitude of the nucleus for this model type.} 
\tablenotetext{c}{The observed quasars whose redshift and nuclear brightnesses correspond to this 
model type.}
\tablenotetext{d}{The average flux lost, in magnitudes, for the models which use the 
given sample of galaxies and the given nuclear brightnesses.} 
\tablenotetext{e}{$M_B$ is calculated from the $H$ flux using spectral index $\alpha$ = $+$0.2}
\end{deluxetable}

\begin{deluxetable}{l c c c }
\tablecaption{Black hole mass and accretion rate estimates \label{bhtab}}
\tabletypesize{\footnotesize}
\tablehead{
\colhead{Name} &\colhead{Redshift} & $M_{BH}/10^{8} M_\odot$\tablenotemark{a} &
$L_{Q}$/$L_{E}$\tablenotemark{b}
}
\startdata
MZZ 9744 &  2.735 & 0.79 & 1.4 \\
MZZ 9592 &  2.710 & 2.19 & 0.7 \\
MZZ 1558 &  1.829 & 1.52 & 0.7 \\
MZZ 11408 & 1.735 & 1.05 & 0.2 \\
MZZ 4935  & 1.876 & 0.29 & 0.7 \\
\enddata
\tablenotetext{a}{The black hole masses derived
from our total $M_V$(host) corrected to $z = 0$, using the van der Marel (1999)
normalization of the $L_{bulge}$--$M_{BH}$ relation. }
\tablenotetext{b}{The ratio of the quasar luminosity to the Eddington
limit, assuming the black hole masses shown here.}

\end{deluxetable}

\clearpage


\end{document}